\documentclass[apj]{emulateapj}

\usepackage{sidecap}
\usepackage{apjfonts}
\usepackage{graphicx}
\usepackage{color}
\usepackage{amsbsy}
\usepackage{mathrsfs}
\usepackage{mathpazo,bm}
\usepackage{amsmath}
\usepackage{multirow}
\usepackage{epstopdf}
\usepackage{amssymb}
\usepackage{xfrac}

\newlength{\hfwidth}
\newlength{\hfwidthsingle}
\renewcommand{\v}[1]{{\boldsymbol{#1}}} 

\newcommand{\del}{\v{\nabla}}
\newcommand{\grad}{\del}

\newcommand{\ttimes}[1]{10^{\tiny{#1}}}
\newcommand{\xtimes}[2]{#1 \times 10^{\tiny{#2}}}

\newcommand{\Figure}[1]{Figure~\ref{#1}}

\newcommand{\Fig}[1]{Fig.~\ref{#1}}
\newcommand{\fig}[1]{\Fig{#1}}
\newcommand{\Figs}[2]{Figs.~\ref{#1} and \ref{#2}}

\definecolor{brown}{rgb}{0.42,0.24,0.07}
\definecolor{darkgreen}{rgb}{0.0,0.6,0.00}
\definecolor{purple}{rgb}{0.7,0.0,0.7}
\definecolor{black}{rgb}{0.0,0.0,0.0}

\shorttitle{Particle Trapping and Streaming Instability in Vortices}
\shortauthors{Raettig et al.}

\begin{document}

\title{Particle Trapping and Streaming Instability in Vortices in Protoplanetary Disks}

\author{Natalie Raettig\altaffilmark{1}, 
        Hubert Klahr\altaffilmark{1}, and
        Wladimir Lyra\altaffilmark{2, 3, 4}}

\altaffiltext{1}{Max-Planck-Institut f\"ur Astronomie, K\"onigstuhl 
                 17, 69117, Heidelberg, Germany;
                 raettig@mpia.de,klahr@mpia.de}
\altaffiltext{2}{Jet Propulsion Laboratory, California Institute of Technology,
                         4800 Oak Grove Drive, Pasadena, CA 91109, USA;
                         wlyra@jpl.nasa.gov,wlyra@caltech.edu}
\altaffiltext{3}{Division of Geological \& Planetary Science,
			California Institute of Technology,
			1200 E California Blvd MC 150-12, Pasadena CA, 91125 USA}               
\altaffiltext{4}{Sagan Fellow}                         

\begin{abstract}
We analyse the concentration of solid particles in vortices created and sustained by radial buoyancy in protoplanetary disks, i.e. baroclinic vortex growth. Besides the gas drag acting on particles we also allow for back-reaction from dust onto the gas. This becomes important when the local dust-to-gas ratio approaches unity. In our 2D, local, shearing sheet simulations we see high concentrations of grains inside the vortices for a broad range of Stokes numbers, ${\rm St}$. An initial dust-to-gas ratio of 1:100 can easily be reversed to 100:1 for ${\rm St}=1$. The increased dust-to-gas ratio triggers the streaming instability, thus counter-intuitively limiting the maximal achievable overdensities. We find that particle trapping inside vortices opens the possibility for gravity-assisted planetesimal formation even for small particles  ($\rm{St}=0.01$) and low initial dust-to-gas ratios (1:$10^4$).
\end{abstract}


\section{Introduction}
A quantitative prescription of planetesimal formation is one of the key issues of planet formation theory.  Models of simple collisional sticking are controversially discussed, both conceptually and also in the explanation of current properties of asteroid and Kuiper belt obejcts \citep{Weidenschilling2011, Morbidellietal2009, Nesvornyetal2011, Bottkeetal2005}. Whether small dust grains stick to one another, bounce or fragment depends on their size and their relative velocities. In general, theory predicts that collisional velocities rise as particles grow, which holds for particles with Stokes number smaller than 1. The Stokes number is defined as ${\rm St} \equiv \varOmega \tau_{\rm f}$, where  $\varOmega$ is the Keplerian frequency, and $\tau_{\rm f}$ is the stropping time of a particle; both drift velocity and turbulence-induced relative velocities have a maximum for St=1, i.e., when the stopping time is of the order of an orbital period \citep[see e.g. review by][]{WeidenschillingCuzzi1993}. At 5\,AU, St=1 corresponds to meter-size objects. Fragmentation occurs at velocities of only a few ${\rm m\,s^{-1}}$, which limits particle sizes to mm-cm \citep{WurmBlum2000,Braueretal2008,BirnstielDullemondBrauer2010}. \cite{Guettleretal2010} and \cite{Zsometal2010} introduced another boundary, the so called bouncing barrier where particles hit one another and bounce without mass transfer. At even smaller size scales, \citet{Okuzumi2011a,Okuzumi2011b} found the charge barrier, where small particles are prevented from approaching one another due to the electric charges built up through collisions with free electrons.
{\citet{Birnstieletal2012} determined the sizes that particles can obtain locally as their growth is limited by radial drift and collisional destructions. They find almost independently on distance from the star a maximum Stokes Number of about $0.01-0.1$ which corresponds to $1-10$ cm at $1$AU, $0.3-3$ cm at $10$ AU and $0.3-3$ mm at $100$ AU \citep[see figs. 5 \& 6 in][]{Birnstieletal2012}.}

To form large planetesimals these difficulties need to be circumvented. One proposed method is gravitational instability \citep{safronov1972, GoldreichWard1973, Johansenetal2006b, Johansenetal2007}: When a sufficient amount of particles is close enough together, their mutual attraction can trigger gravitational collapse, rapidly forming large planetesimals that then sweep up small particles \citep{LambrechtsJohansen2012}. Different methods to capture dust have been studied, such as zonal flows by \citet{Johansenetal2011} and \citet{Dittrichetal2013}, pressure rings around stars \citep{Whipple1972, KlahrLin2001, KlahrLin2005}, convection cells \citep{KlahrHenning1997} or vortices either numerically \citep{Tangaetal1996, Johansenetal2004, Lyraetal2008b, Lyraetal2009a, Lyraetal2009b, Meheutetal2012a,Meheutetal2012b} or analytically \citep{BargeSommeria1995, Chavanis2000, KlahrBodenheimer2006, ChangOishi2010, LyraLin2013}.

From these studies it has become clear that dust can easily concentrate in anti-cyclonic vortices. However, feedback of dust on the vortical flow, which can potentially frustrate particle concentration, has not been studied in detail. This is the topic of the present study. 

Besides \citet{Lyraetal2008b,Lyraetal2009a,Lyraetal2009b,Meheutetal2012b,Ataieetal2013,ZhuStone14}, and \citet{Zhuetal14}, who analyzed particle trapping in vortices excited by the Rossby wave instability, other works did not choose a particular vortex formation mechanism. In our study, vortices are naturally produced by the radial stratification of disks \citep{KlahrBodenheimer2003, PetersenJulienStewart2007,PetersenStewartJulien2007, LesurPapaloizou2010, LyraKlahr2011, Raettigetal2013}. The growth of the vortices occurs proportionally to the radial buoyancy frequency (aka Brunt-V\"ais\"al\"a frequency) squared $N^2$  \citep{Raettigetal2013} and the local thermal relaxation time scale as argued by \citet{LesurPapaloizou2010}, numerically confirmed by \citet{Raettigetal2013} .
The buoyancy frequency, which is a function of the relative scale height $H/r$, logarithmic radial entropy and pressure gradients $\beta_K, \beta_P$ and the adiabatic index $\gamma$:
\begin{equation}
N^2 = - \frac{1}{\gamma}\left(\frac{H}{r}\right)^2 \beta_K \beta_P.
\end{equation}
{A linear theory to explain this behaviour has been put forward by \citet{KlahrHubbard2014} who identify the instability as a radial "convective overstability" in accretion disks. \citet{Lyra2014} performed 3D simulations of this "convective overstability" and showed that in the non-linear phase indeed vortices were emerging from the flow.}

In this paper we set out to assess how efficiently dust can be concentrated in vortices enforced by realistic values for the radial stratification in temperature and density, also with a plausible value for the thermal relaxation time. We do this as a first step via two-dimensional, local simulations. 
Ultimately, only threedimensional stratified runs will be able to include all relevant physics from vortex stability \citep{BarrancoMarcus2005, LesurPapaloizou2009, LyraKlahr2011} to sedimentation of dust. Nevertheless, here we find even in 2D that particle concentration in the vortices is sufficiently intense that back-reaction of the particles onto the gas motion has to be considered. We find that the streaming instability \citep{YoudinGoodman2005, Johansenetal2006b} is triggered, counter-intuitively limiting the maximum dust-to-gas ratio, and severely perturbing the gas flow inside the vortex. In some cases we see vortices getting disrupted and later reforming, starting a new cycle of particle concentration.

In simulations similar to ours but three-dimensional and more numerically expensive, where particles were trapped in zonal flows of magneto-hydrodynamical origin and including self-gravity, the overdensities  that we found in the present paper did already lead to gravitational collapse \citep{Johansenetal2006a, Johansenetal2007} and to the formation of planetesimals. On the other hand 2D simulations including particle feedback are ideally suited to study a wide parameter range to learn whether streaming instability can be triggered and how it affects vortex stability.


The paper is structured as follows. We first review the underlying physics of dust-motion in a gas disk, specifically in a vortex including dust-gas interactions. Then we describe the numerical setup. The results of our simulations are given in Section \ref{2Dresults}. Here we especially look at the reached dust-to-gas ratios compared to the average initial dust-to-gas ratio. {In Section 5 we discuss the prospects to form planetesimals via gravitational fragmentation of the dust enhancements in vortices and in Section 6 we present our measurements of collisional speeds among dust grains.} Finally, we summarize and conclude in Section \ref{2Dconclusion}.

\section{Physical Background}
{The evolution of the gas component of the disk in the Pencil Code is given by the Navier Stokes Equation containing stellar gravity as well as the virtual forces of the rotating and shearing box as explicit terms \citep{LyraKlahr2011}.
Dust grains on the other hand are evolved by solving the equations of motions for Lagrangian particles (see Section 3).

In general the gas and dust feel the same external forces except the pressure force $f_p=-\rho_{\rm g}^{-1}\nabla p$, where $\rho_{\rm g}$ is the gas density and $p$ the pressure. This term only affects the gas. For instance the global pressure gradient in the gas leads to a sub-Keplerian orbital gas velocity $\v u$.}The corresponding buoyancy force for a particle is $f_{p,\rm s}=-\rho_{\rm s}^{-1}\nabla p$, where $\rho_{\rm s}$ is the material density of the solid material, can be neglected because $\rho_{\rm s} \gg \rho_{\rm g}$. Since the particles do not feel this global pressure, they need to orbit at Keplerian velocities in order to be in centrifugal balance with stellar gravity. The resulting velocity difference between gas and large particles acts as a headwind on the particles which decreases their angular momentum leading to radial drift inward. Smaller particles get dragged along by the gas and therefor feel a net acceleration towards the star also making them drift inward.

The time on which the dust particles adjust to the gas velocity is the friction time. {For subsonic Epstein drag this is $\tau_s={\rho_{\rm s} s}( {\rho_{\rm g} c_s})^{-1}$ \citep{Weidenschilling1977b} which depends on particle size $s$ and local sound speed $c_s$. The Epstein regime considers particles smaller than the mean free path of gas. Larger particles have to be treated in the Stokes regime. But as long as relative velocities between dust and gas are small, it is possible to describe the coupling force to be linear with respect to the relative velocity $f =  m \frac{\delta u}{\tau_s}$ independent on the detailed calculation of $\tau_s$. Especially it is not necessary to determine the shape and density of the particles or the gas density and sound speed until one asks for the particle size corresponding to a given friction time.

This particle gas coupling is usually expressed in terms of the dimensionless Stokes number ${\rm St}\equiv\varOmega\tau_s$. Particles of different size, shape and density, but ultimately with the same ${\rm St}$ will behave the same as far as aerodynamics are concerned.}

{The friction time and thus the Stokes number is a function of the local gas density. We nevertheless assume the Stokes number to be constant, which is justified as the gas density fluctuates only by a few percent in the sub-sonic turbulence we consider in our  simulations. }

Since our simulations are only two-dimensional we can not consider vertical settling of particles.
{The fact that the real 3D gas disk is vertically stratified and thus particles of a given size have a vertically varying friction time, has to be discussed in more detail. One can interpret this approximation as focussing on a narrow vertical range around the midplane, in which neither the gas nor the particle stratification are too strong. Then one can assume that also the friction time and the dust to gas ratio has no strong vertical variations. This assumption is a little less stringent than saying that all particles are actually in the midplane.} 

{The general behavior of particle drift in a non-laminar yet geostrophic flow\footnote{In a geostrophic flow the pressure gradients are in equilibrium with the centrifugal and Coriolis acceleration.} can be understood easily from a a simplified
treatment of the Lagrangian equations of motion for gas and dust:

\begin{eqnarray}
\label{udot}\partial_t{ \v u}&= &\v F-\frac{1}{\rho_{\rm g}}\grad p - \frac{\rho_{\rm d}}{\rho_{\rm g} \tau_s}\left(\v u -\v v\right)\\
\label{vdot}\partial_t{ \v v}&=& \v F-\frac{1}{\tau_s}\left(\v v-\v u\right),
\end{eqnarray}
where $\v v$ is the particle velocity and $\v F$ collects the terms that are the same for both dust and gas, e.g., gravitational force from the central star. The second term of the equations describe the pressure force, and the third term is the friction force between gas and dust particles. With some math it can be shown that by subtracting Eq.\ (2) and (3) and assuming $|\partial_t{ \v u}-\partial_t{ \v v}|\ll |{ \v u} - { \v v}| / {\tau_s}$ particles will move towards regions with higher pressure ${ \v v} = { \v u} + {\tau_s} \grad p$
even if the pressure maximum is tiny and the profile relatively smooth \citep[see e.g.][]{KlahrLin2001}.}

The vortices we consider are anti-cyclonic vortices and have a slightly lower epicyclic frequency than the Keplerian orbital frequency. Since there is a geostrophic balance between the Coriolis and the pressure force a high pressure region inside the vortex is created. Particle accumulation inside a vortex then basically works the same way as the radial drift  in an accretion disk works. The particles do not feel the pressure gradient inside the vortex and therefore their epicyclic frequency equals the Keplerian frequency. Yet, the rotation frequency of a pressure supported vortex is smaller than the epicyclic frequency. Thus the headwind from the gas causes the particles to lose eccentricity and forces the particles to  spiral towards the center of the vortex. For an in depth analysis of vortex capturing including a comparison of all forces acting on the gas and dust inside the vortex see \citet{AdamsWatkins1995, BargeSommeria1995, Tangaetal1996, Chavanis2000} and \citet{Johansenetal2004}.

If the dust density becomes comparable to the gas density the drag forces from particles onto the gas can no longer be neglected as long as the St is smaller than $\approx 10$. This back-reaction can alter the motion of the gas and also leads to even higher dust concentrations through the streaming instability \citep{YoudinGoodman2005}. {The last term of equation (\ref{udot}) represents the back-reaction from dust grains onto the gas.}

\begin{table}
\caption[]{Simulation setup and $\varepsilon_{\rm max}$}
\label{setup}
\begin{center}
\begin{tabular}{l c c c c c c}\hline
run & ${\rm St}$ & $\sfrac{\Sigma_\mathrm{d,0}}{\Sigma_\mathrm{g,0}}$ & \tiny{feedback} & $\varepsilon_{\rm max}$&\tiny{Fraction of $M_{\rm d}$} & \tiny{Fraction of $M_{\rm d}$}\\ 
& & & & &\tiny{with $\varepsilon \ge 0.01$(\%)} & \tiny{with $\varepsilon \ge 1$(\%)}\\\hline
NF1  & 0.01& $1:100$	   & no &6.26	  & 86.64	& 2.88\\
NF2	& 0.05  & $1:100$  	   & no & 636.00& 92.31	& 75.63\\
NF3	& 1  & $1:100$  	   & no & 843.70& 79.71	& 1.26\\
NF4  & 20& $1:100$	   & no & 46.81 & 99.90	& 99.80\\
F1 	& 0.01  & $1:100$  	    & yes& 1.07 & 70.40	& 0\\
F2	& 0.05  & $1:100$  	    & yes& 3.86 & 95.24	& 2.13\\
F3	& 1  & $1:100$  	    & yes& 77.33& 98.73	& 83.74\\
F4 	& 20  & $1:100$  	    & yes&0.73	& 79.82	& 8.09\\
DG1 & 0.05& $1:1\, 000$    & yes& 1.15& 86.74	& 0\\
DG2 & 0.05& $1:10\, 000$  & yes& 0.70& 78.30	& 0\\
DG3 & 1  & $1:1\,000$    & yes& 11.53& 98.46	& 56.37\\
DG4 & 1  & $1:10\, 000$ & yes& 4.17	& 97.30	& 76.15\\\hline
\end{tabular}
\end{center}
\end{table}


\begin{table}
\caption[]{Collisional velocities}
\label{v_coll}
\begin{center}
\begin{tabular}{ c c   c c c c c c c}\hline
%
${\rm St}$ & $\varepsilon_0$   &$\alpha$& $\frac{v_{\rm coll}}{c_s}$& $\frac{u_{\rm rms}}{\sqrt{\alpha} c_s}$& $\frac{v_{\rm rms}}{\sqrt{\alpha} c_s}$& $\frac{v_{\rm coll}}{\sqrt{\alpha} c_s}$& $\frac{\Delta v_{\rm coll}}{\sqrt{\alpha} c_s}$&$\frac{v_{\rm coll, OC}}{\sqrt{\alpha} c_s}$\\
& & \tiny{($/{\ttimes{-2}}$)} & \tiny{($/{\ttimes{-3}}$)}& & & \tiny{($/{\ttimes{-2}}$)} &  \tiny{($/{\ttimes{-2}}$)} & \\\hline
1   & $\ttimes{-2}$ &1.03 &   2.77 &2.84 &1.37 &2.73 &2.56 & 1\\
1   & $\ttimes{-3}$ &1.56 &   3.17 &4.36 &3.61 &2.54 &3.34 &1 \\
1   & $\ttimes{-4}$ &1.54 &   1.25 &3.88 &2.27 &1.01 &4.01 &1\\
0.05 & $\ttimes{-2}$ & 1.56 &  2.86 &3.27 &2.41 &2.29 &0.59 &0.31\\
0.05 & $\ttimes{-3}$ & 1.57 &  2.71 &4.29 &3.80 &2.16 &1.28 &0.31\\
0.05 & $\ttimes{-4}$ & 1.54 &  2.08 &3.88 &3.04 &1.68 &0.09 &0.31 \\\hline
\label{v_coll}
\end{tabular}
\end{center}
\end{table}

\section{Numerical setup}\label{2DNumSetup}

We perform two dimensional shearing sheet simulations with the {\sc Pencil Code} {where we consider the vertically integrated densities $\Sigma$ instead of the three-dimensional densities $\rho$.} The Euler-equations for the gas are solved on a Cartesian grid, identical to the setup of \citep{Raettigetal2013}, but now augmented by a term for the particle feedback on the gas $\varepsilon\left(\v u -\v v\right)/\tau_s$ with the dust-to-gas ratio $\varepsilon = {\Sigma_{\rm d}}/{\Sigma_{\rm g} }$
\begin{eqnarray}
\frac{\mathcal{D}\Sigma_{\rm g}}{\mathcal{D}t} &+&\left(\v{u} \cdot\grad\right)\Sigma_{\rm g}=-\Sigma_{\rm g}\grad\cdot\v{u}+ f_D(\Sigma_{\rm g})\label{PSrho},\\
\label{NS}\frac{\mathcal{D}\v{u}}{\mathcal{D}t}&+&\left(\v{u} \cdot\grad\right)\v{u}=-\frac{1}{\Sigma_{\rm g}}\grad p - 2\varOmega_0\left(\v{\hat{z}}\times\v{u}\right)\nonumber\\
&+& \frac{3}{2}\varOmega_0u_x\v{\hat{y}}+\frac{\beta p_0}{R_0}\left(\frac{1}{\Sigma_{\rm g}}-\frac{1}{\Sigma_0}\right)\v{\hat{x}}+f_\nu(\v{u},\Sigma)\label{PSu}\nonumber\\
&-& \frac{\Sigma_{\rm d}}{\Sigma_{\rm g} \tau_s}\left(\v u -\v v\right),\\
\frac{\mathcal{D}{s}}{\mathcal{D}t}&+&\left(\v{u} \cdot\grad\right){s}=\frac{1}{\Sigma_{\rm g} T} \bigg\{\grad \cdot \left(K\grad T\right)-\Sigma_{\rm g} c_v\frac{(T-T_0)}{\tau_{\rm cool}} \nonumber\\
&+& \frac{\beta p_0}{r_0}\frac{u_x}{\left(\gamma -1\right)}\bigg\}+f_K(s)\label{PSs}.
\end{eqnarray}
Here, $\Sigma_{\rm g}$ is the gas density, $\v{u}$ is the deviation of the gas velocity from the Keplerian value, $s$ the entropy, $T$ the temperature, $c_v$ the specific heat at constant volume, $\tau_{\rm cool}$ is the thermal relaxation time scale, and $K$ the heat conductivity. The symbol
\begin{equation}
\frac{\mathcal{D}}{\mathcal{D}t}\equiv \frac{\partial}{\partial t}+u_y^{\left(0\right)}\frac{\partial}{\partial y}
\end{equation}
represents the Keplerian derivative where $u_y^{\left(0\right)}=-3/2\varOmega_0x$ 
{and $\varOmega_0$ is the Keplerian frequency at the orbit of the shearing box $R_0$.  The radial deviation from $R_0$ is given by $x$ and the linearized azimuthal direction is now measured in terms of $y$. 
Boundary conditions are periodic in $y$ and shear-periodic in $x$. Further terms in the equations are diffusion terms to ensure numerical stability of the finite difference Pencil Code $f_D(\Sigma_{\rm g}), f_\nu(\v{u},\Sigma)\label{PSu}, f_K(s)\label{PSs}$ and radial stratification of the disk in terms of pressure and entropy for radially constant density: $\beta = -\frac{{\rm d} {\log} p}{{\rm d} {\log} R} = -\frac{{\rm d} {\log} s}{{\rm d} {\log} R}$. The stratification term $\beta$ occurs in the linearized component of radial buoyancy (Eq. (5)) and in the term for radial transport of entropy (Eq. (6)) as derived in \citet{LyraKlahr2011}. These terms in combination with thermal relaxation are the driver of vortices as found in \citet{PetersenJulienStewart2007}}.
{As described in their paper gas that is moving radially outward is generally warmer and of lower density than gas moving inward. The density mismatch across the vortex leads to a mismatch in buoyancy and thus a vortex will feel a torque accelerating it. The maximal amplification occurs when thermal relaxation is on the same order as the internal rotation period of the vortex.}

The dust grains are modeled with a particle approach. For each individual particle we solve the equation of motion including gas drag. We do not allow for self-gravity so far. In principle, for an individual particle it does not matter if there are other particles in the simulation or not. However, in the simulations where we allow for back-reaction of the particles on the gas, there is an indirect influence of one particle on the other particles due to the altered gas velocity.

The gas disk is co-rotating with Keplerian velocity at the co-rotational radius $r_0$. But for simulations involving both dust and gas we need to include a velocity offset due to the pressure gradient. The pressure gradient is balanced by the Coriolis force which is the linearized expression for the radial centrifugal acceleration in the local co-rotating system
\begin{equation}
2\varOmega u_y=\frac{1}{\Sigma_{\rm g}}\frac{\partial p}{\partial r}=c_s\varOmega\frac{H}{r}\frac{\partial\ln p}{\partial\ln r}.
\end{equation}
The deviation by which the gas velocity is lower than the Keplerian velocity $u_{\rm K}=\varOmega r$ is \citep{Nakagawaetal1986}
\begin{equation}
\eta=0.5\left(\frac{H}{r}\right)^2\frac{\partial\ln p}{\partial\ln r},
\end{equation}
which leads to a sub-Keplerian gas velocity and a deviation from the Keplerian velocity $u_{\rm K}$ of 
\begin{equation}
u_y=0.5\left(\frac{H}{r}\right)^2\frac{\partial\ln p}{\partial\ln r}\varOmega r=\eta u_{\rm K}.
\end{equation}
This velocity deviation $\eta$ is added to the simulations artificially. 

For each particle $i$ the equation of motion is solved individually via

\begin{equation}
\frac{d \v v^{\left(i\right)}_*}{dt} = 2\varOmega v_{*y}^{\left(i\right)}\v{\hat{x}} - \frac{1}{2}\varOmega v_{*x}^{\left(i\right)}\v{\hat{y}} - 
       \frac{1}{\tau_{\rm s}}\left(\v v_*^{\left(i\right)} - \v u \left(\v x ^{\left(i\right)}\right)\right),
\end{equation}
where $\v v^{\left(i\right)}_* = \v v  - \eta u_{\rm K}$ is the particle velocity corrected by the velocity offset and $\v u \left(\v x ^{\left(i\right)}\right)$ it the gas velocity at the position of the dust particle. The index $*$ is omitted from now on.

For our 2D simulations we choose $\eta=-0.01$ which corresponds to a pressure gradient of $\beta=2$ and a disk aspect ratio of $h=0.1$.
The terminal velocities of the particles for a given Stokes number in a laminar disk are now in terms of $\eta$ \citep{Weidenschilling1977b}
\begin{eqnarray}
v_x&=&\frac{2\eta u_{\rm K}}{{\rm St +{St}^{-1}}}\label{rad_drift},\\
v_y&=&\frac{\eta u_{\rm K}}{{\rm St^2}}.
\end{eqnarray}
These are the best assumptions one can do for the initial velocities of particles in our simulations.

Our physical domain spans 4 disk scale-heights, $\pm2H$, around the co-rotational radius in radial-direction ($x$-axis) and $16H$ in azimuthal-direction ($y$-axis). The grid itself consists of $288^2$ grid cells\footnote{The number of 288 grid cells comes from the architecture of our computational cluster which needs a multiple of 12 to parallelize the {\sc Pencil Code} efficiently.}. We choose $\beta=2$ for the entropy and pressure gradients which is a relatively strong gradient compared to gradients we expect in protoplanetary accretion disks are between $\beta=0.5$ and $\beta=1$. However, \citet{Raettigetal2013} found that the general behavior of vortices is the same for weak and strong entropy gradients. The development of the vortices is merely faster with stronger gradients and we can scan a large parameter space with a reasonable amount of computing time. For a first estimation the strong gradient is sufficient, but in our future 3D studies including self-gravity we will use $\beta= 1$ because then we aim for quantitative results on planetesimal formation rates and mass spectrum. 

We first evolve the disk for $200$ local orbits without particles. This way we make sure that a fully grown, long lived vortex has developed before we put in particles. Physically, this corresponds to various scenarios: A) a vortex developing in the outer parts of the disk and then migrating inwards through regions with particles, B) the growth of particles into a size regime where they get trapped by the nearest vortex or C) a radial flux of particles close to bouncing and / or drift barrier that enters the vicinity of the vortex.

After the initial 200 orbits, we distribute $400\, 000$ particles randomly in the disk. This corresponds to 4-5 particles per grid cell initially. That way we minimize numerical effects that can arise if there are not enough particles in the computational domain, e.g. the effect a single particle has can be extremely overestimated if not enough particles are considered. Whenever we refer to times in this paper, we mean time elapsed since particles were put into the simulation.
{At the time when the dust to gas ratio approaches unity and streaming instability is triggered, 500 or more particles are in one grid cell. Following \citet{JohansenYoudin2007} these numbers are safely beyond the minimum particle number per grid cell needed to achieve numerical convergence.}

{Starting our simulations with particles included would have little changed our results as the initial and average dust to gas ratios are much too low in order for the dust to have impact on the gas dynamics right from the beginning. Retrospectively this assumption was later justified in simulations (see Section 4.2) in which the vortex gets disrupted via the streaming instability and then forms again, this time in the presence of dust grains. This second growth phase does not differ from the initial growth phase without dust grains present.}

Each of the $400\, 000$ particles represents one super-particle, a collection of several particles, all of the same size, and a given mass according to the initial dust-to-gas ratio.

Generally we have an initial dust-to-gas ratio $\varepsilon_0$ of 1:100 which means that the disk consists out of 1\% solid material and 99\% gas. Note that in two dimensions the {\sc Pencil Code} assumes surface densities instead of volume densities. Therefore the dust-to-gas ratios we talk about in this paper refer to $\varepsilon=\Sigma_{\rm d}/\Sigma_{\rm g}$ rather than to $\varepsilon=\rho_{\rm d}/\rho_{\rm g}$.
To simulate different particle sizes we use different Stokes numbers of $\rm{St}=0.01$, $\rm{St}=0.05$, $\rm{St}=1$ and $\rm{St}=20$. 
{By only defining the Stokes Number we are able to cover particles both in the Epstein as well as in the Stokes regime as friction forces are linear with respect to velocity.}
At 5 AU this corresponds to particles between $3\, \mathrm{mm}$ and $6\, \mathrm{m}$ for $\Sigma_{\rm g}=300\, {\rm g\, cm^{-2}}$ and $\rho_{\rm s}=2\, {\rm g \, cm^{-3}}$ . The parameters for all our simulations can be seen in Table \ref{setup}. A separate simulation is carried out for each ${\rm St}$.

{Particles growth due to collisions is not treated in our simulation. Physically particles would have a growth time of several thousand years, therefore we neglect this effect for the sake of keeping the simulation simple and easy to be evaluated.}

\section{Results}\label{2Dresults}

\begin{figure}
\resizebox{\columnwidth}{!}{\includegraphics{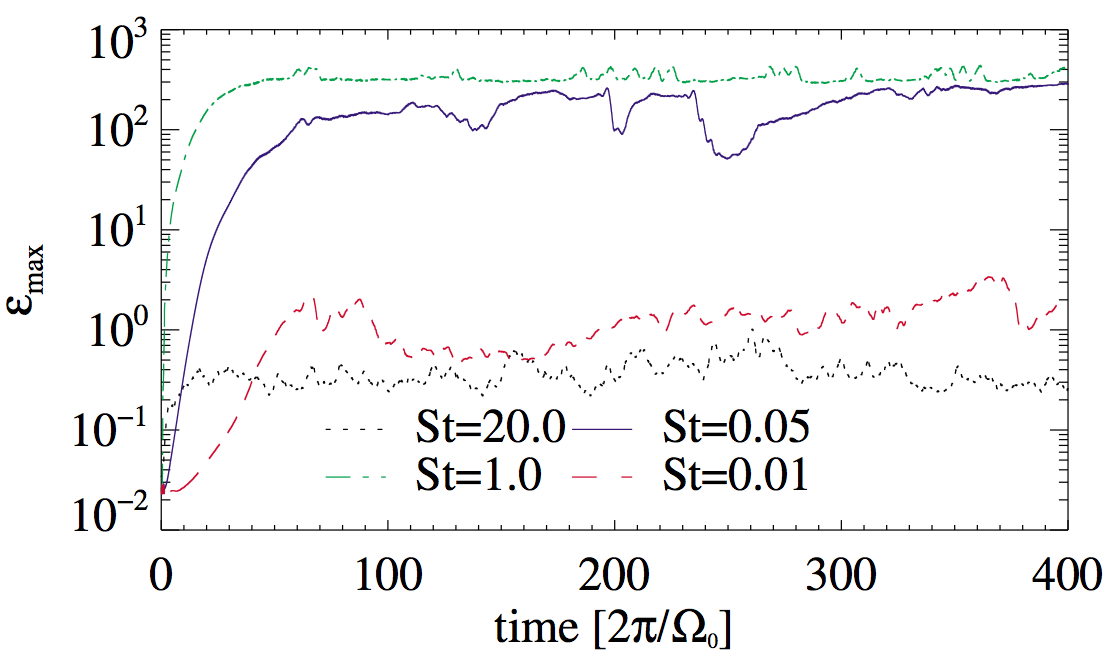}}
\caption[$\varepsilon_{\rm max}$ for different ${\rm St}$ without particle feedback]{Maximum dust-to-gas ratio, $\varepsilon_{\rm max}$, for simulations without particle feedback. The different lines represent the different particle sizes: dotted (black) line: $\rm{St}=20$, dash-dotted (green) line: $\rm{St}=1$, solid (blue) line: $\rm{St}=0.05$, and dashed (red) line: $\rm{St}=0.01$. Almost all particles of intermediate size ($\rm{St}=1$ and $\rm{St}=0.05$) accumulate in the vortex. Strongly coupled particles ($\rm{St}=0.01$) accumulate only partially, because they also couple to the gas outside of the vortex. Large particles ($\rm{St}=20$) hardly couple to the gas at all and therefore are not affected by the vortical motion.}
\label{edg-nfb}
\end{figure}

\begin{figure}
\resizebox{\columnwidth}{!}{\includegraphics{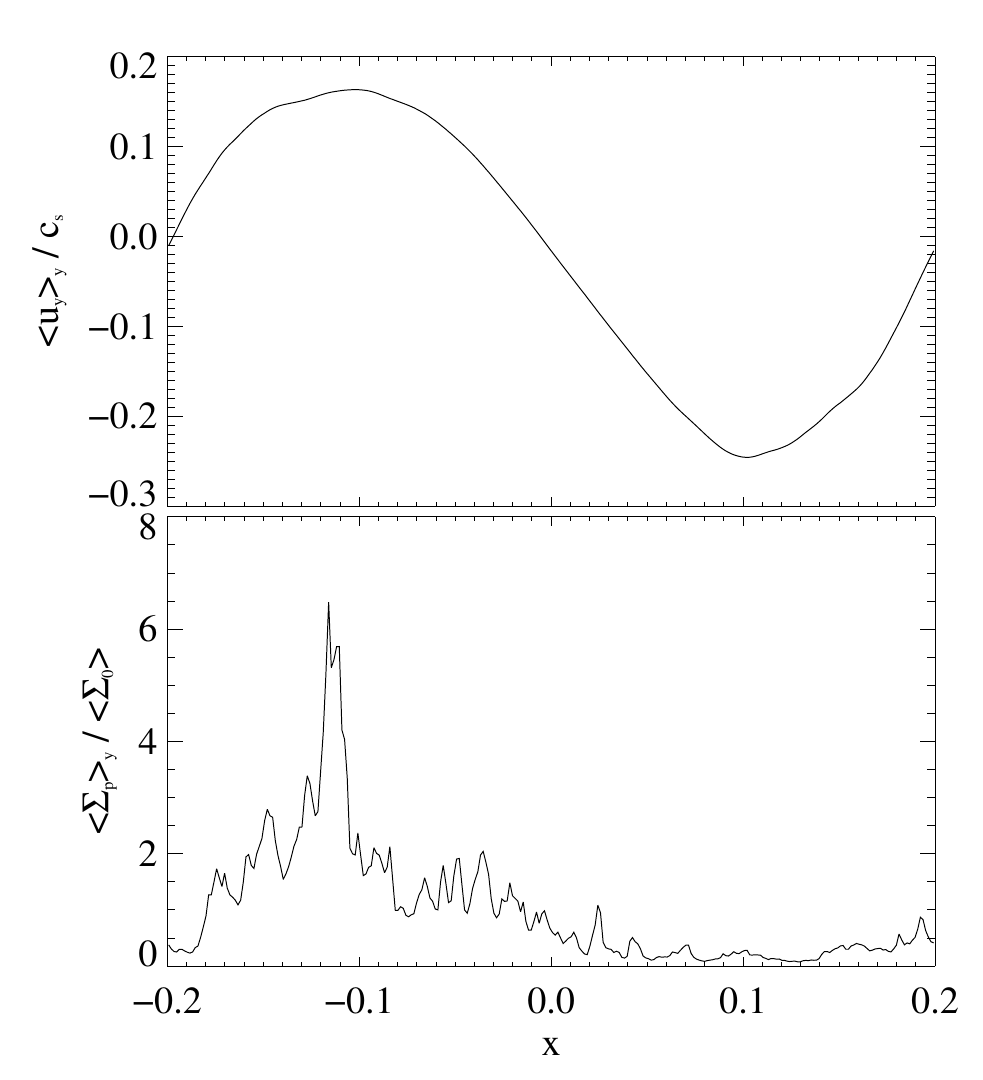}}
\caption{Azimuthally averaged gas velocity and particle density for St=20 particles. The particles accumulate in the zonal flow structure. }
\label{zonal_flow}
\end{figure}

\begin{figure*}
\centering
\resizebox{0.425\textwidth}{!}{\includegraphics{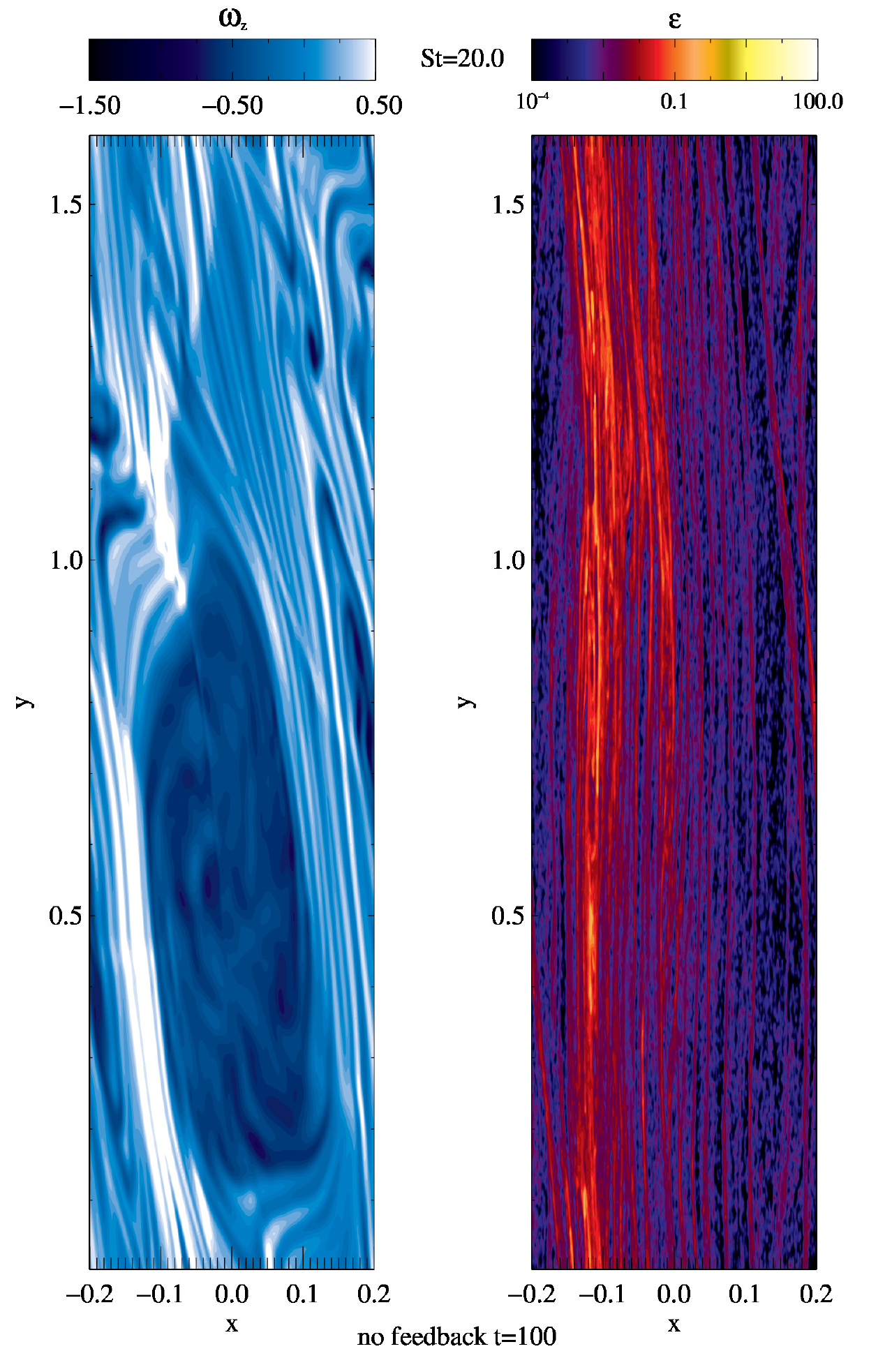}}
\resizebox{0.425\textwidth}{!}{\includegraphics{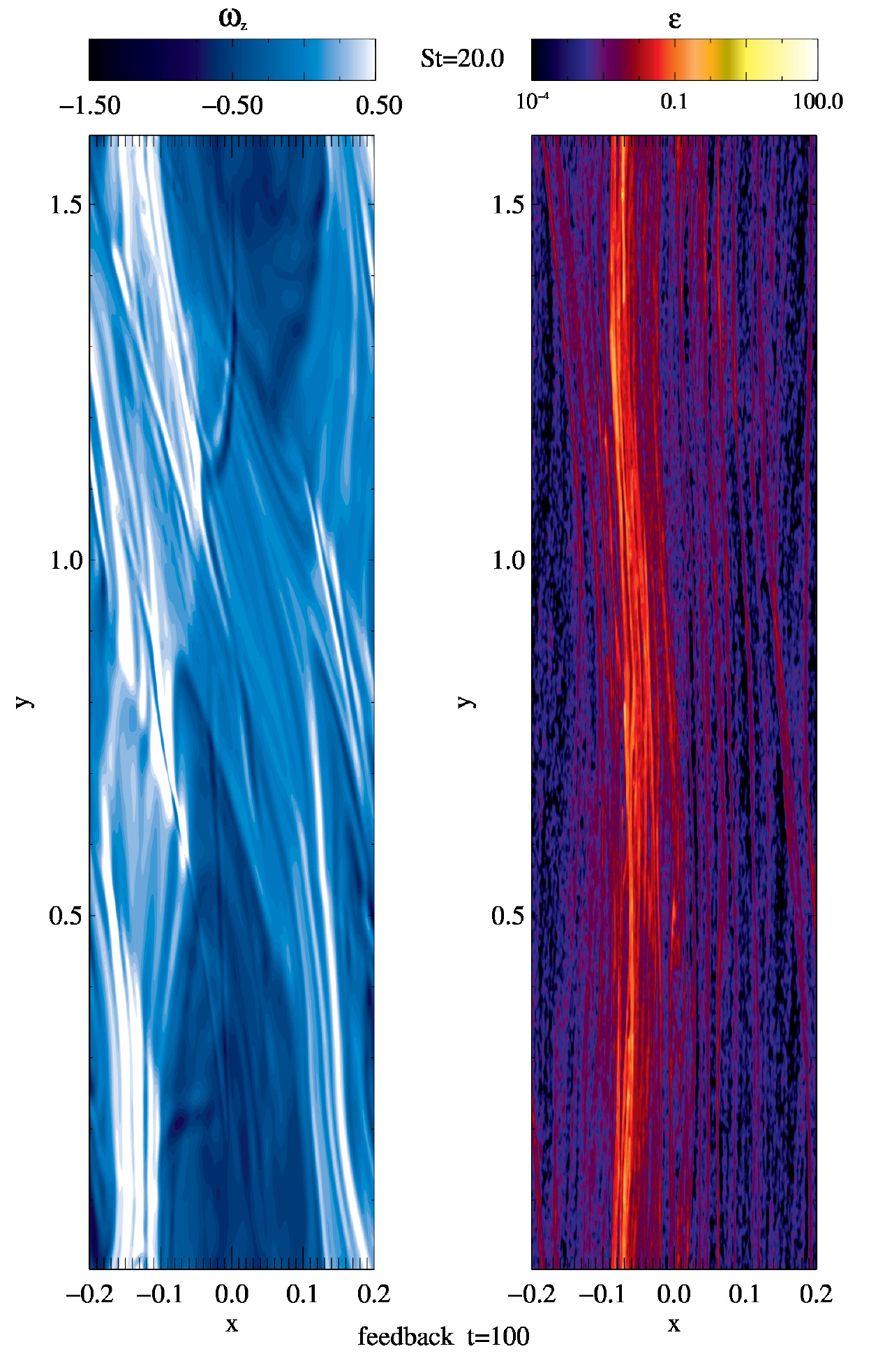}}
\resizebox{0.425\textwidth}{!}{\includegraphics{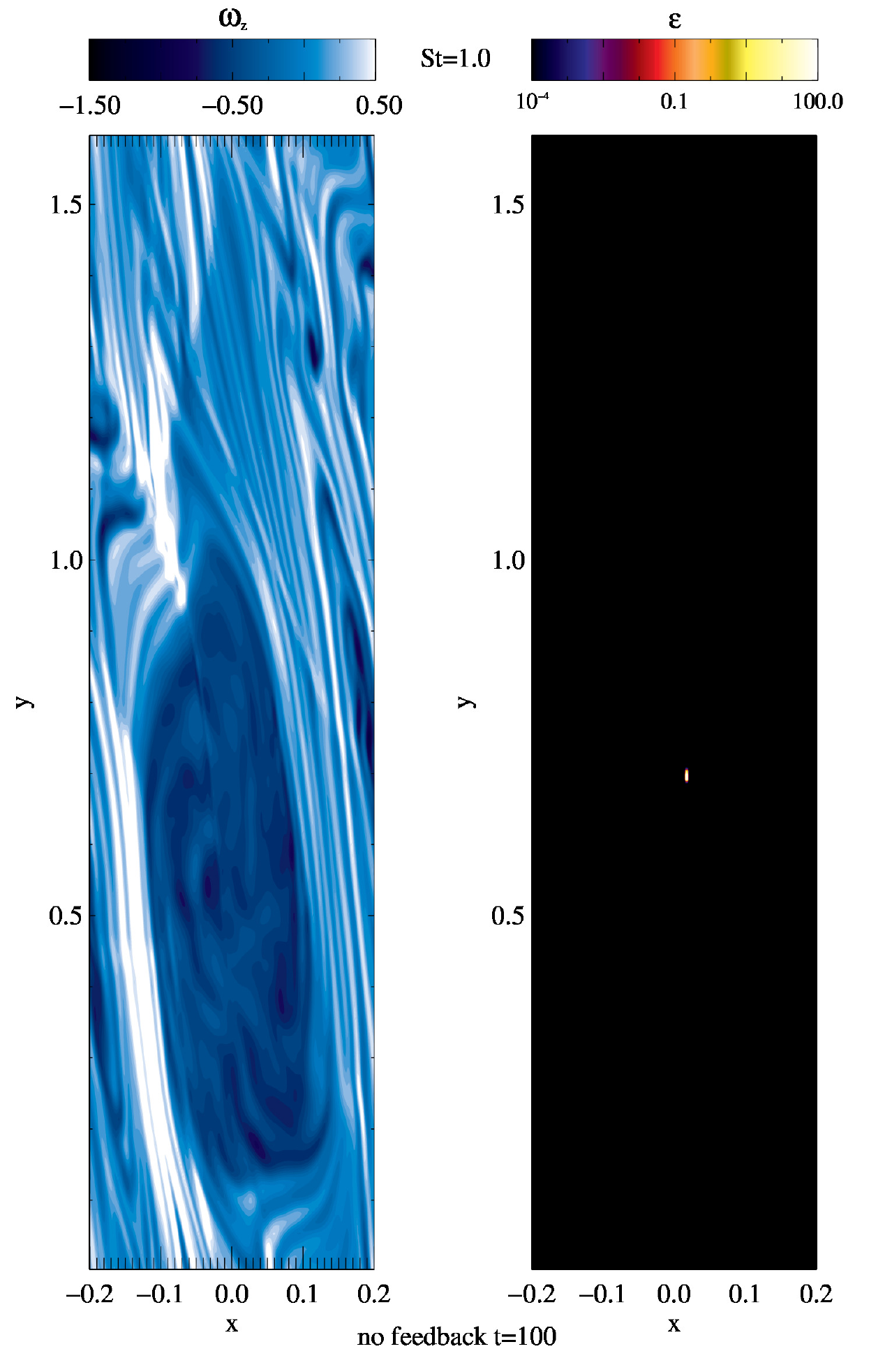}}
\resizebox{0.425\textwidth}{!}{\includegraphics{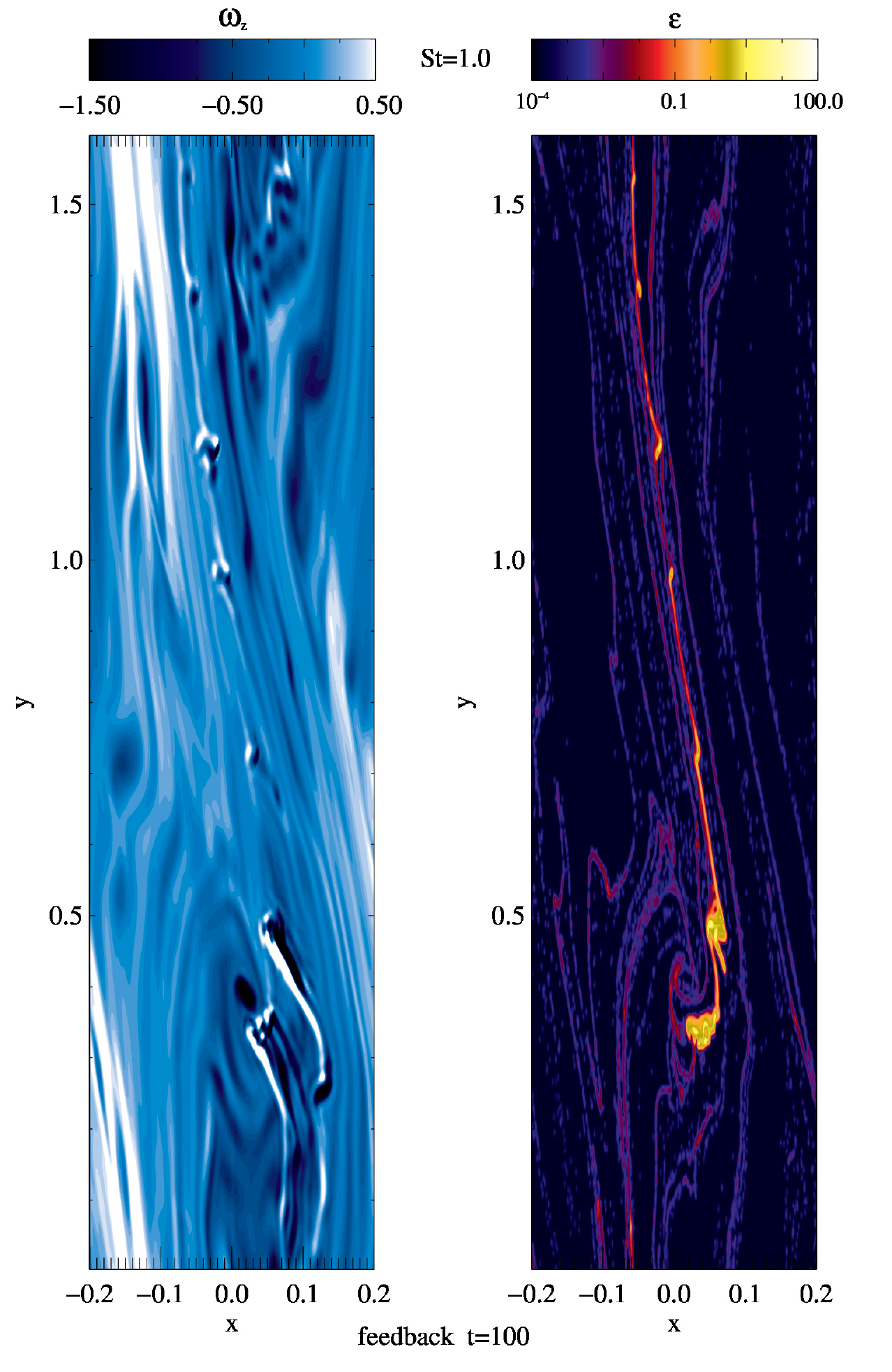}}
%
\caption[Comparison between simulations with and without feedback. $\rm{St}=20$ and $\rm{St}=1$]{Each of the four panels show vorticity $\omega_z$ (in units of the local Keplerian frequency $\varOmega_0$), and dust-to-gas ratio $\varepsilon$. The left-hand-side panels do not include particle feedback, whereas the ones in the right-hand-side do. The particle sizes correpond to Stokes numbers $\rm{St}=20$ (top) and $\rm{St}=1$ (bottom).}
\label{St20u1}
\end{figure*}

\begin{figure*}
\centering
\resizebox{0.425\textwidth}{!}{\includegraphics{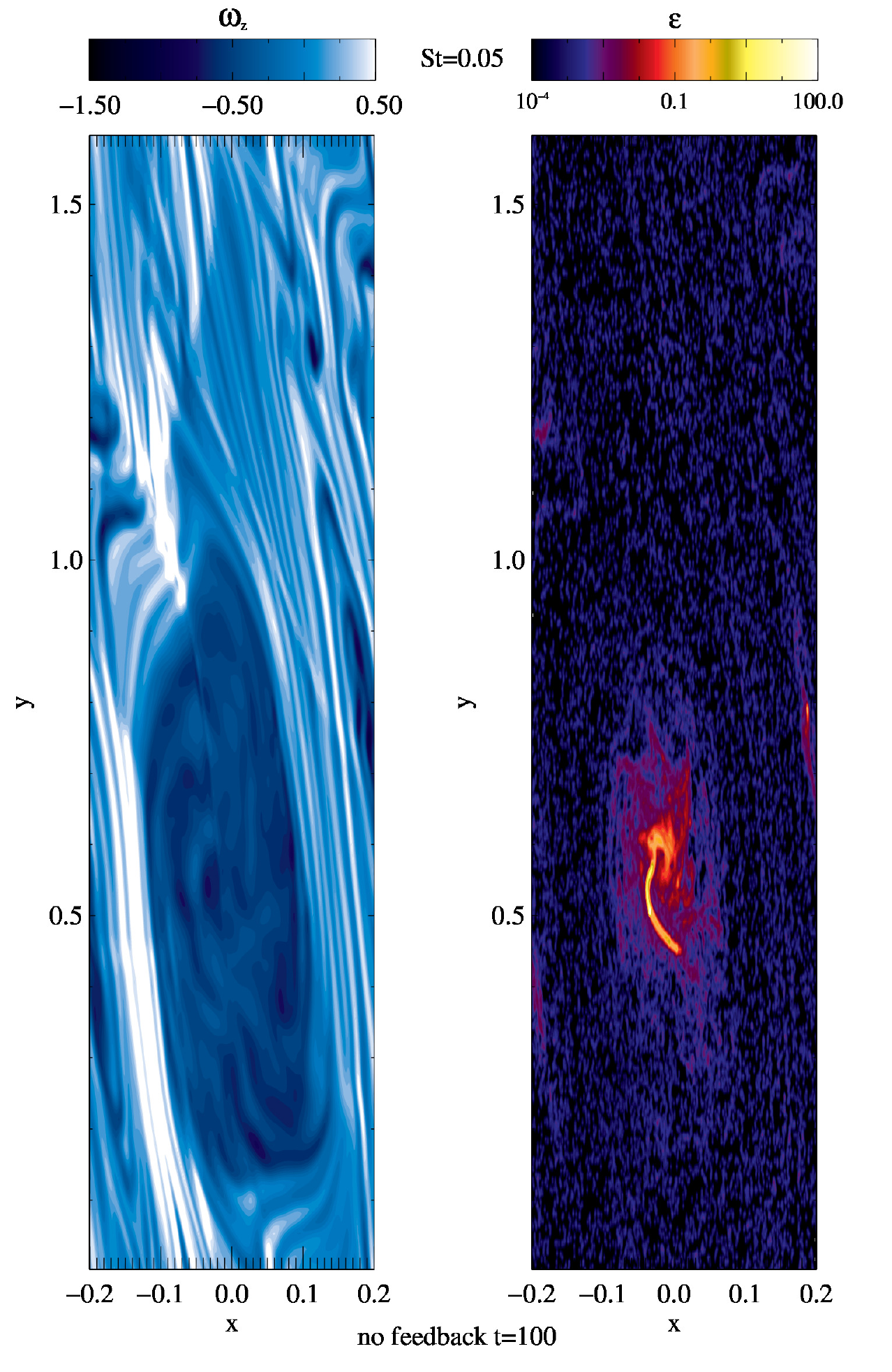}}
\resizebox{0.425\textwidth}{!}{\includegraphics{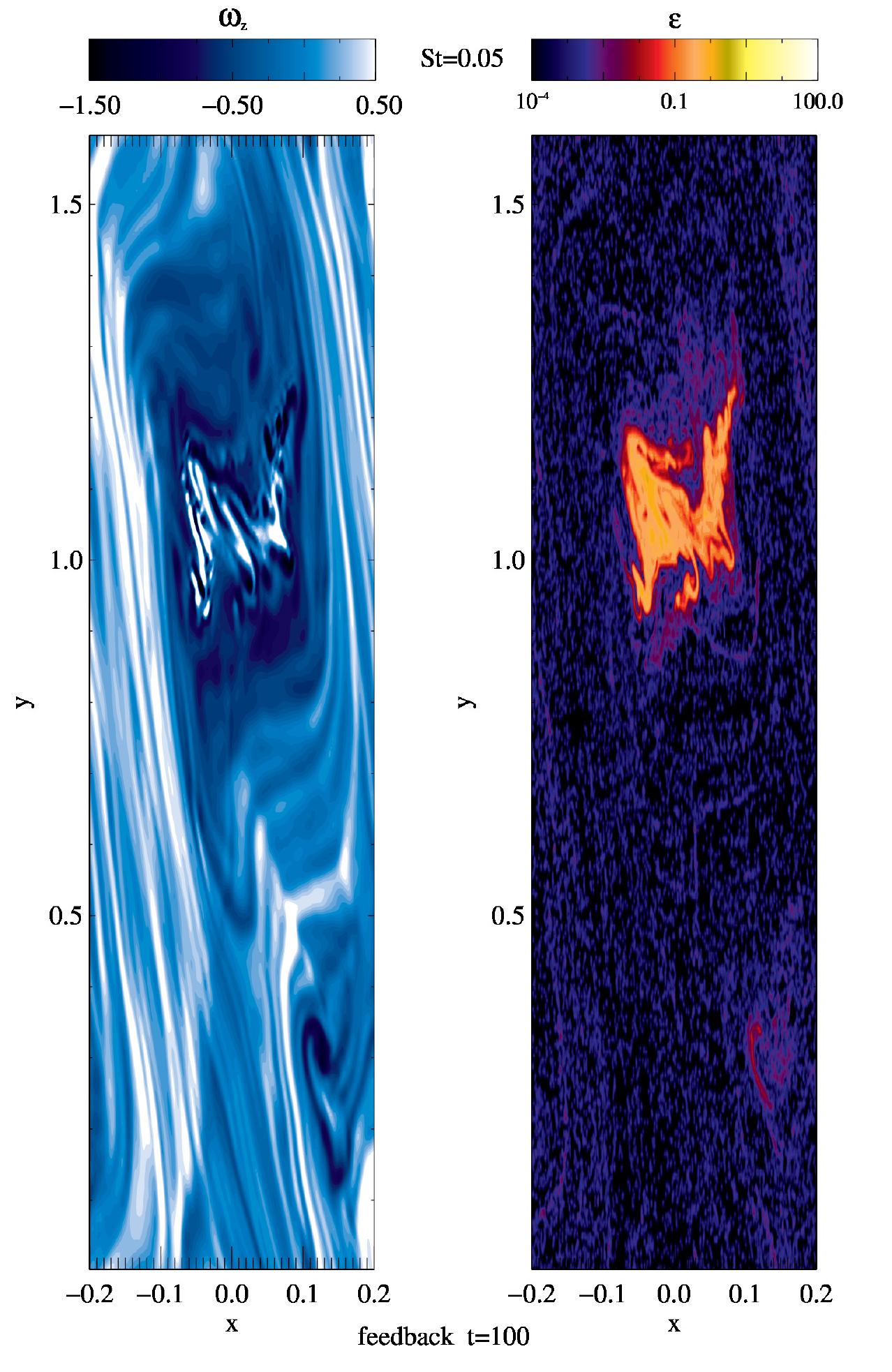}}
\resizebox{0.425\textwidth}{!}{\includegraphics{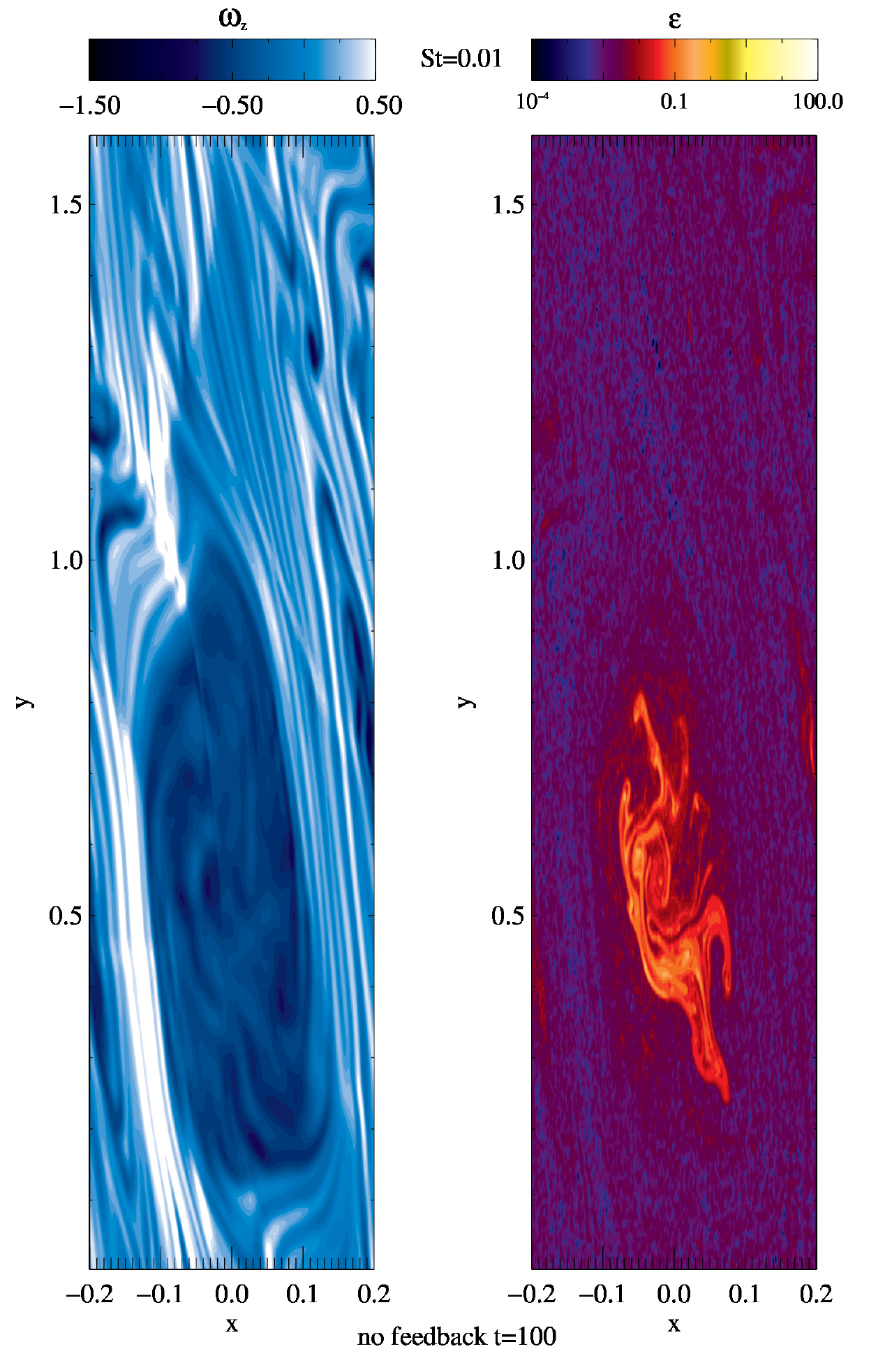}}
\resizebox{0.425\textwidth}{!}{\includegraphics{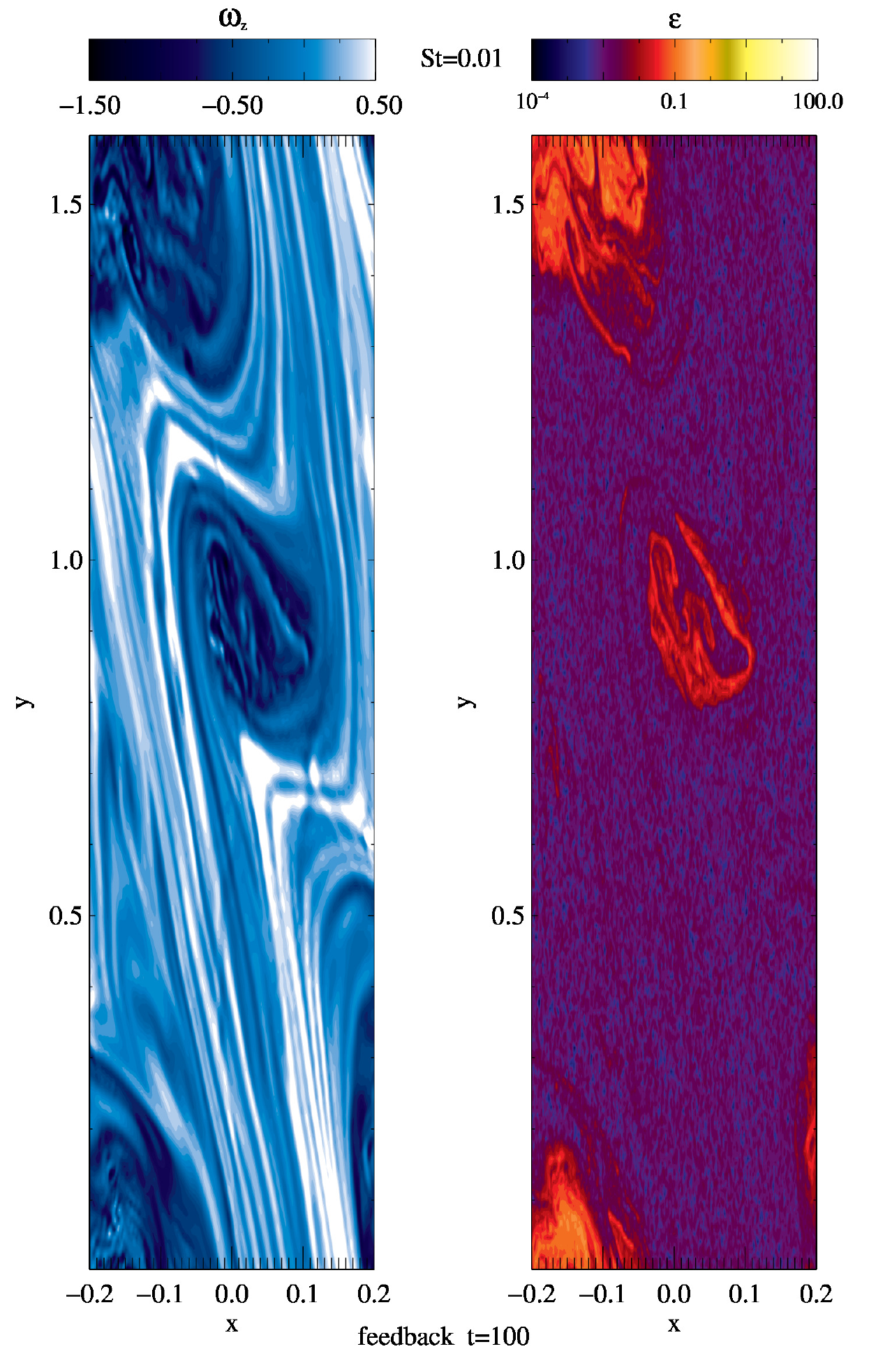}}
%
\caption[Comparison between simulations with and without feedback. $\rm{St}=0.05$ and $\rm{St}=0.01$]{Same as \fig{St20u1}, but for particles of $\rm{St}=0.05$ (top) and $\rm{St}=0.01$ (bottom).}
\label{St05u01}
\end{figure*}

\begin{figure}
\centering
\resizebox{\columnwidth}{!}{\includegraphics{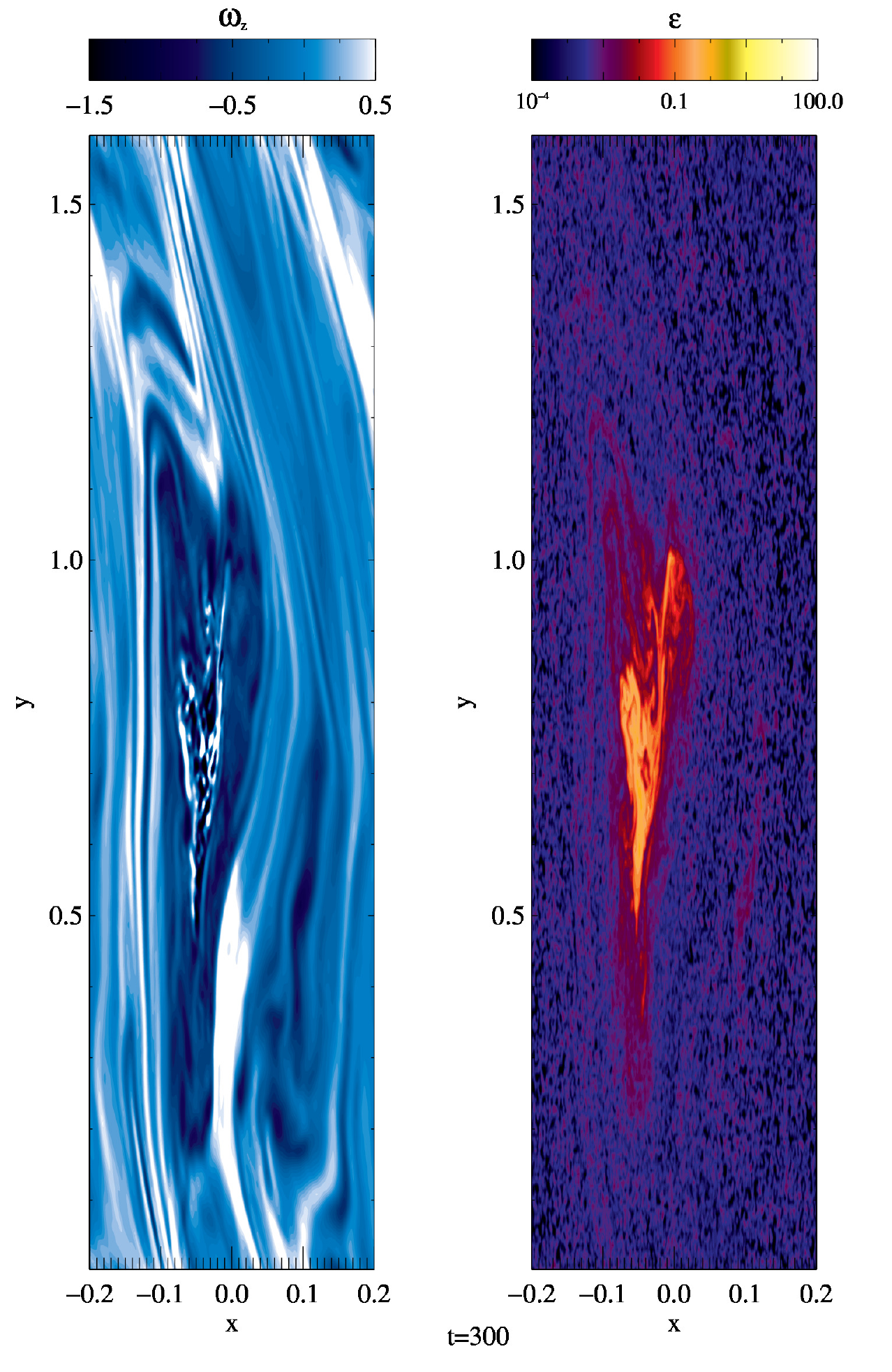}}
\caption[$\omega_z$ in units of the local Keplerian frequency $\varOmega_0$and $\varepsilon$ for $\rm{St}=0.01$ including feedack]{Gas vorticity (left panel) and dust-to-gas ratio (right panel) for a simulation with $\rm{St}=0.01$ particles and back-reaction onto the gas. The elliptical vortical gas flow is distinguishable in the vorticity plot. There are strong accumulations of particles within the vortex. Although many particles spread out over the entire vortex, most particles concentrate in the center of the vortex. The positive vorticity values in the vortex (light areas in the left plot) show the effect that particles have on the gas. Where the gas encounters obstacles, namely high particle concentrations, steep vorticity gradients develop.}
\label{St01_sigma}
\end{figure}

\begin{figure*}
\centering
\resizebox{0.425\textwidth}{!}{\includegraphics{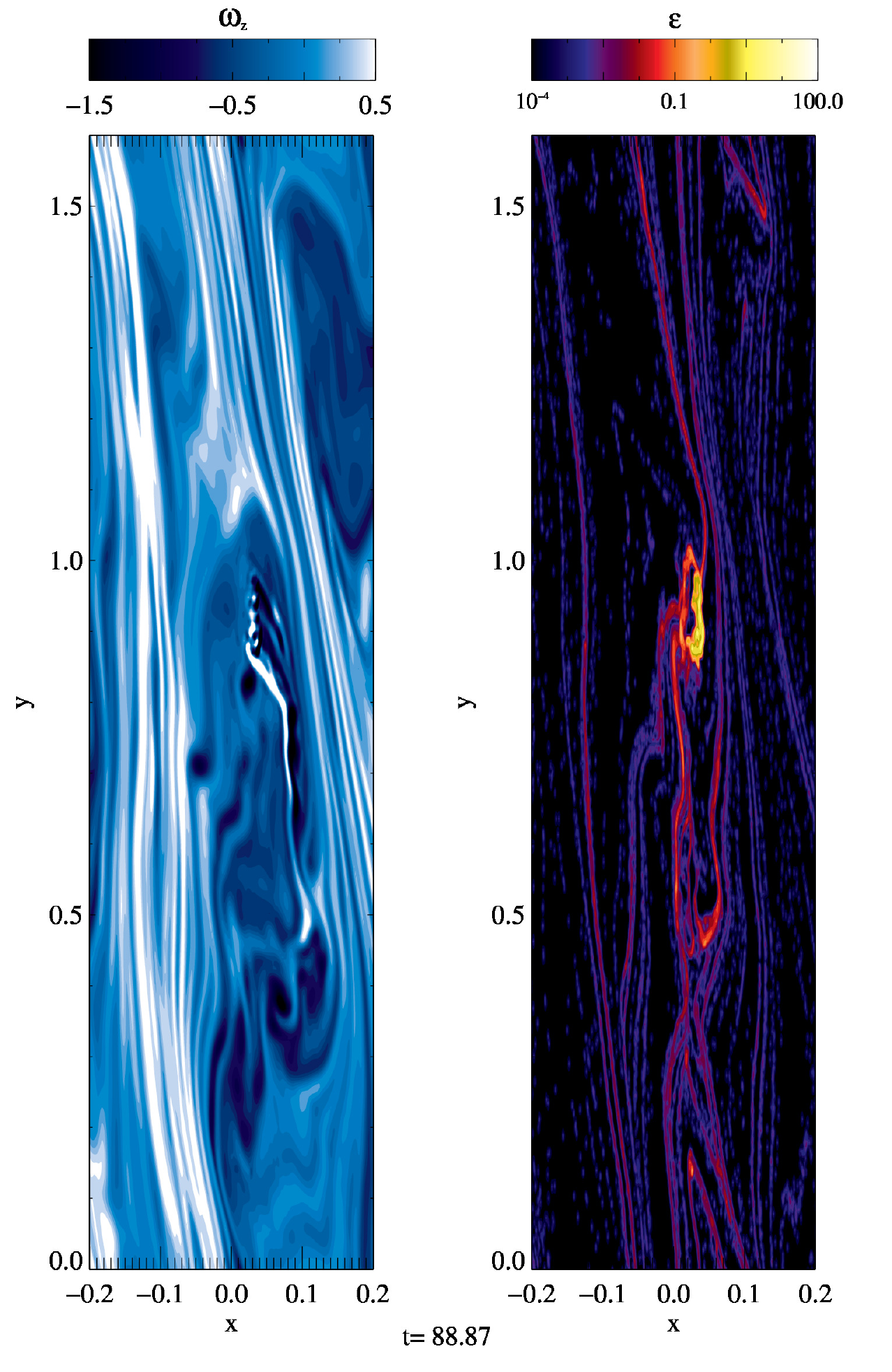}}
\resizebox{0.425\textwidth}{!}{\includegraphics{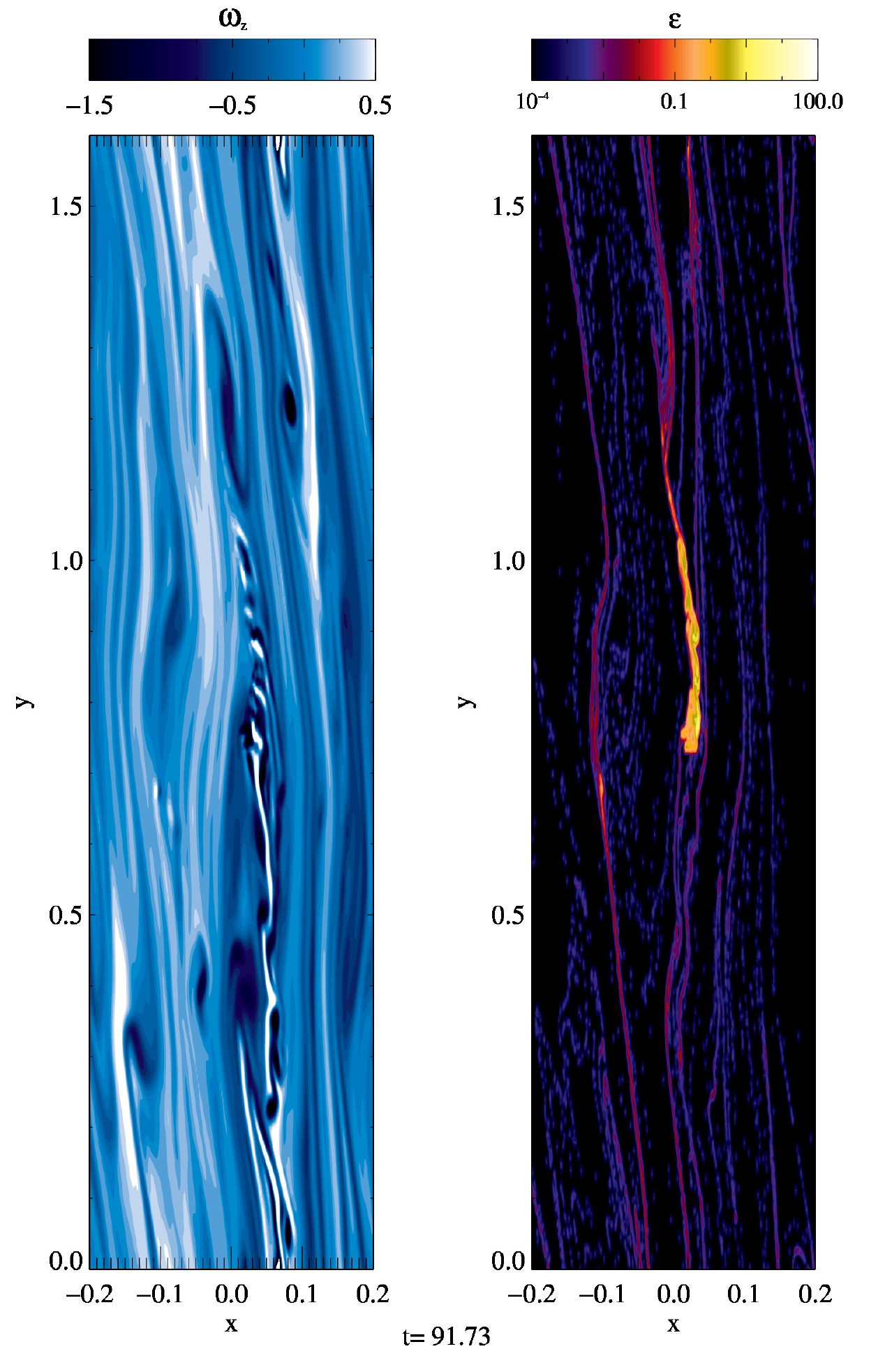}}
\resizebox{0.425\textwidth}{!}{\includegraphics{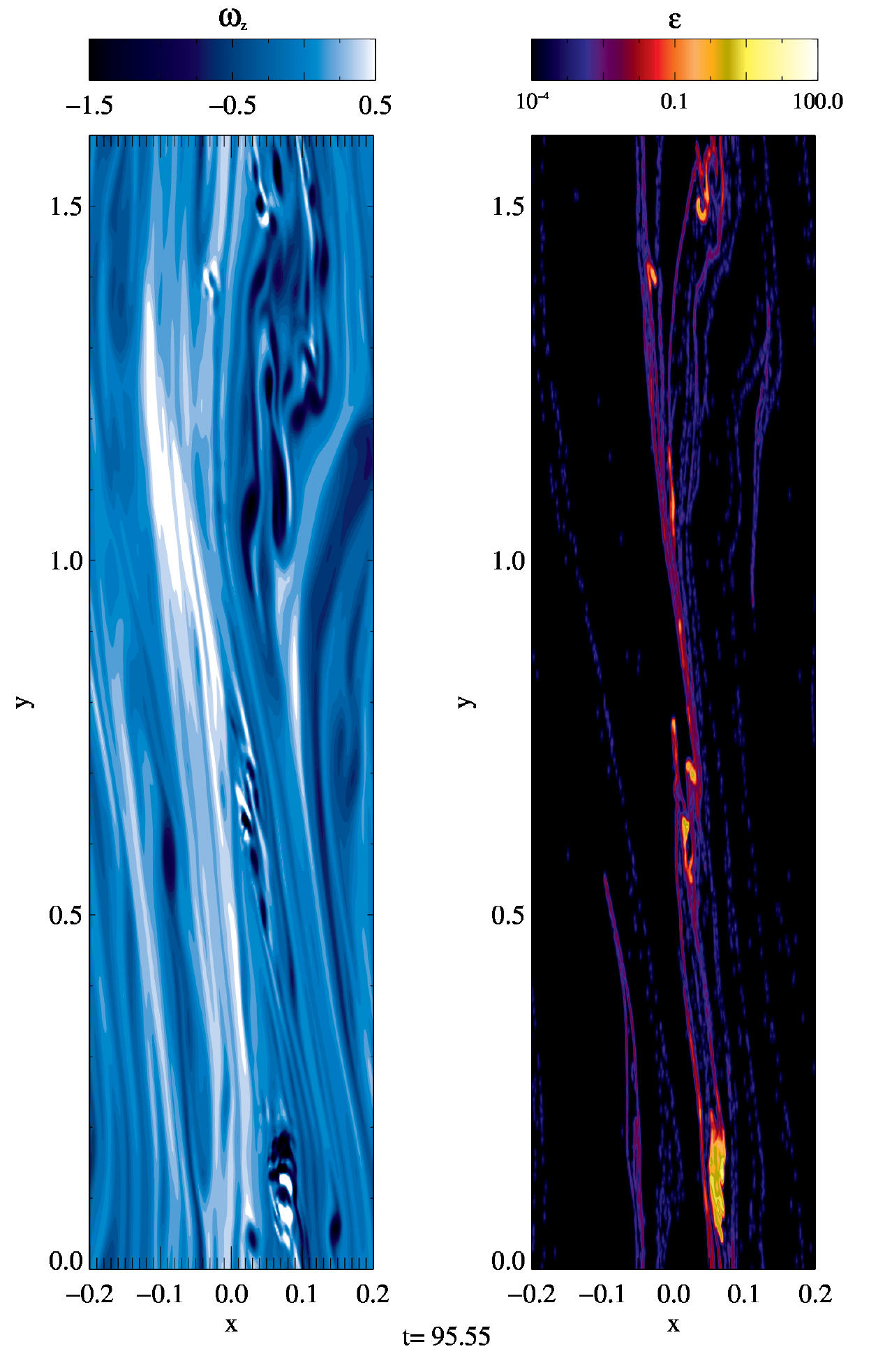}}
\resizebox{0.425\textwidth}{!}{\includegraphics{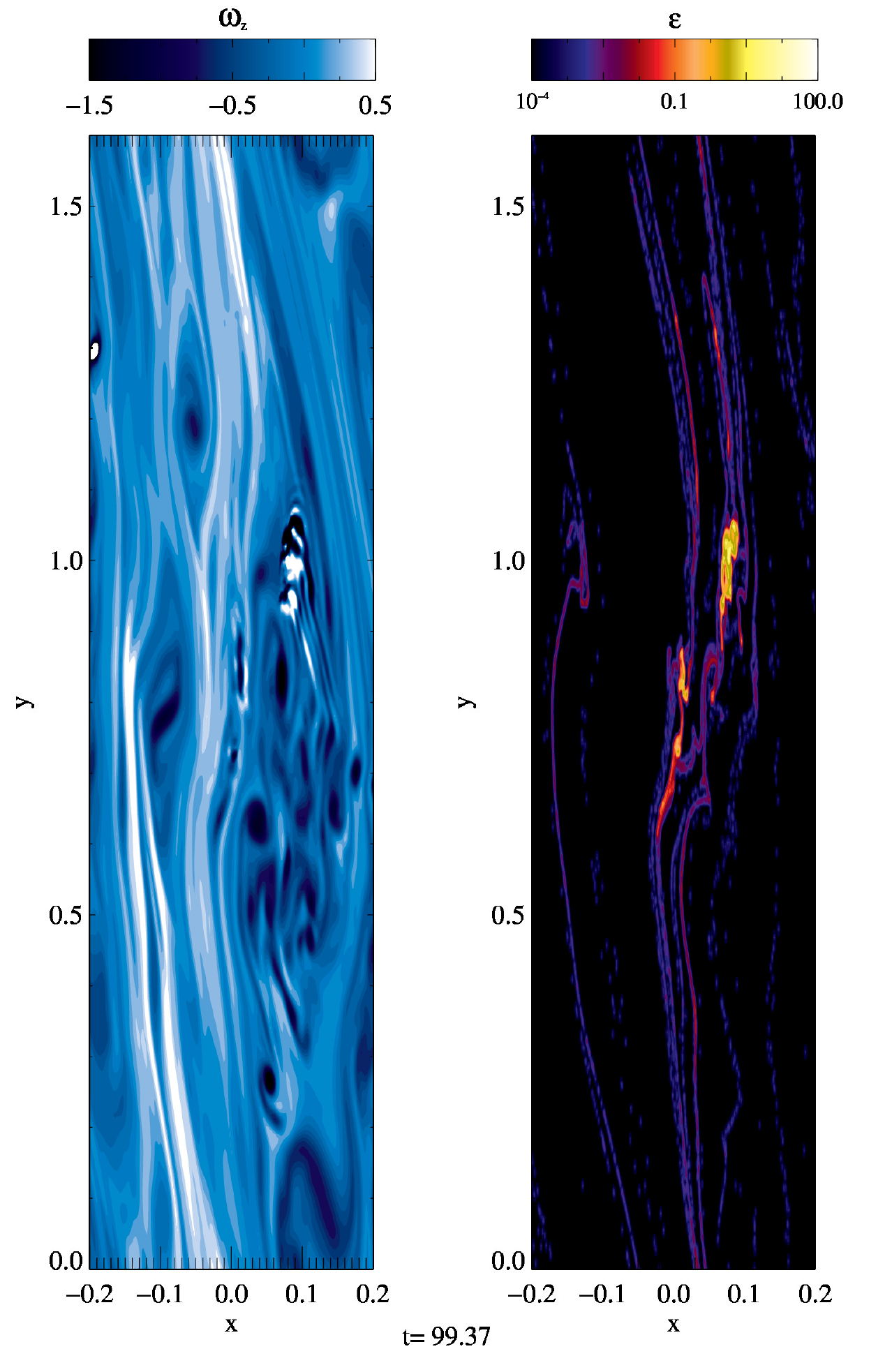}}
%
%
\caption[Destruction of vortex due to particle concentration of $\rm{St}=1$]{Vorticity $\omega_z$ in units of the local Keplerian frequency $\varOmega_0$ (1st and 3rd column) and dust-to-gas ratio (2nd and 4th column) for $\rm{St}=1$ particles for different points in time. At the first snapshot, the vortex is still clearly distinguishable. In the second snapshot, it has been disrupted strongly by the particle accumulation. As this particle accumulation spreads out, the vortex can slowly regain its shape (3rd snapshot) and form a large, yet still perturbed, vortex again (4th snapshot).}
\label{snapshots}
\end{figure*}

\begin{figure}
\resizebox{\columnwidth}{!}{\includegraphics{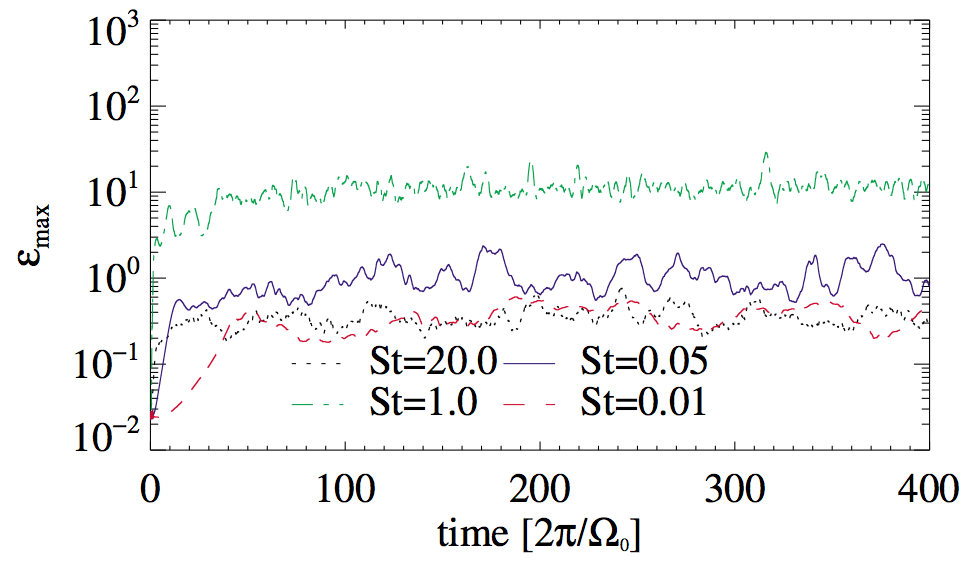}}
\caption[$\varepsilon_{\rm max}$ for different ${\rm St}$ including particle feedback]{Maximum dust-to-gas ratio, $\varepsilon_{\max}$, for simulations with particle feedback. The different lines represent the different particle sizes: dotted (black) line: $\rm{St}=20$, dash-dotted (green) line: $\rm{St}=1$, solid (blue) line: $\rm{St}=0.05$, and dashed (red) line: $\rm{St}=0.01$. Particles of $\rm{St}=1$ reach the highest dust enhancements and concentrate in a very local area in the vortex, while smaller particles spread out over the entire vortex. Therefore the overdensities reached are lower.}
\label{edg_fb}
\end{figure}

\begin{figure}
\resizebox{\columnwidth}{!}{\includegraphics{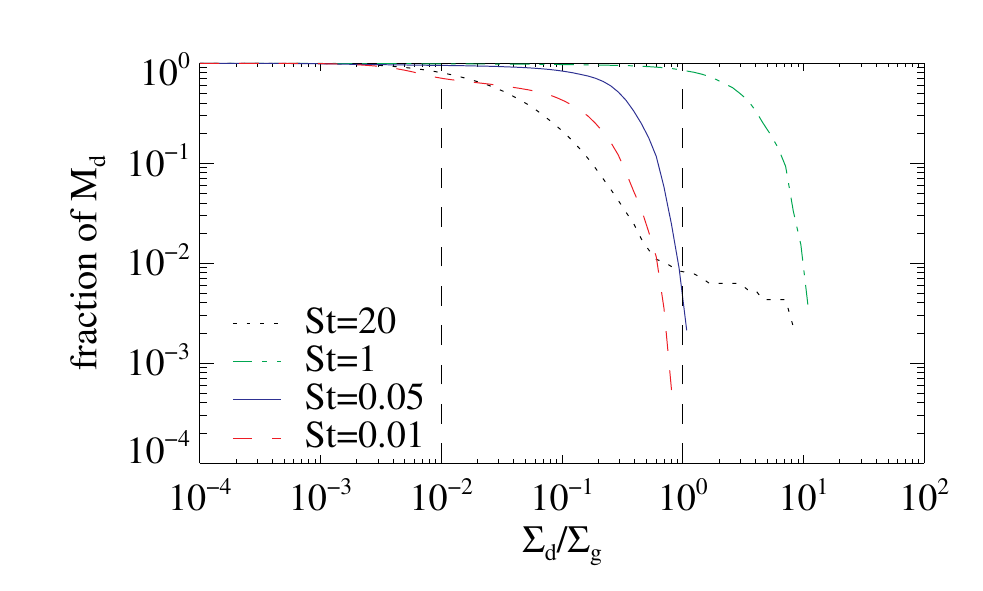}}
\caption[Particle clustering]{Fraction of the entire dust mass in a specific dust-to-gas ratio or higher. $\varepsilon \ge 1$ are reached for all particle sizes. For $\rm{St}=1$ particles more than 80\% of all particles have concentrated in areas with larger $\varepsilon$ than 1.}
\label{clustering}
\end{figure}

\begin{figure}
\resizebox{\columnwidth}{!}{\includegraphics{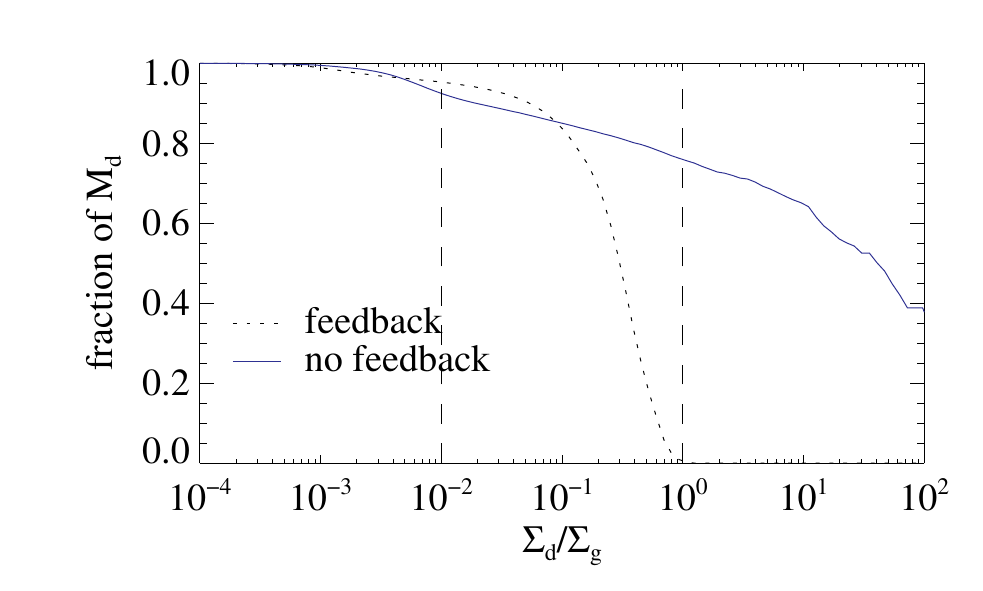}}
\caption[Particle clustering]{Fraction of the entire dust mass in a specific dust-to-gas ratio or higher. $\varepsilon \ge 1$ are reached for St=0.05 particles with (dashed black line) and without feedback (solid blue line).}
\label{compare}
\end{figure}

\begin{figure}
\resizebox{\columnwidth}{!}{\includegraphics{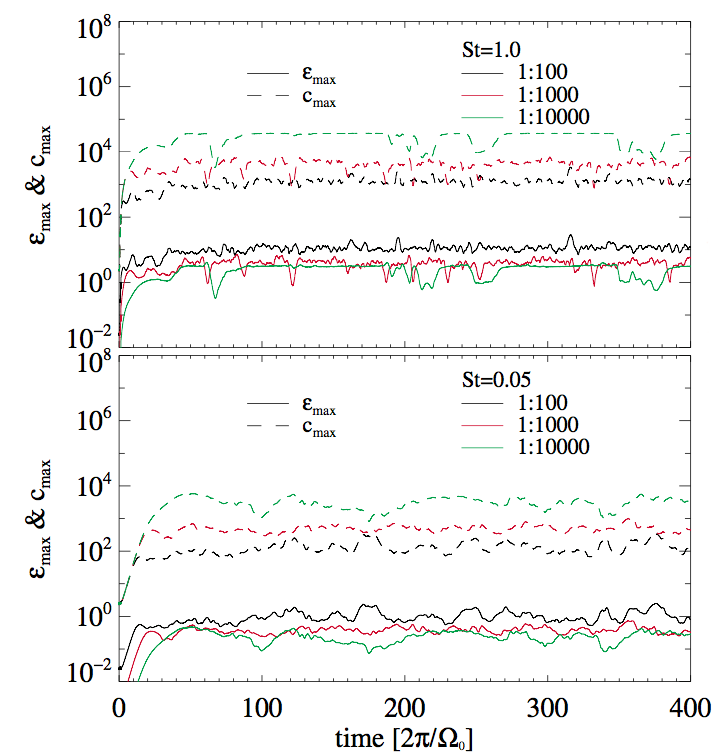}}
\caption[Particle concentration and $\varepsilon$ for different $\varepsilon_0$ and $\rm{St}=1$ and $0.05$]{Particle concentration (solid lines) and maximum dust-to-gas ratios for $\rm{St}=1$ (top) and $\rm{St}=0.05$ (bottom) particles. The color represent the different initial dust-to-gas ratios: $1:100$ (black), $1:1\,000$ (red), and $1:10\,000$ (green). More individual super-particles are captured in the vortices for a low initial $\varepsilon$, because their back-reaction is less efficient. Since each of these super-particles is less massive than with higher $\varepsilon$, the overall dust-to-gas ratio for low initial $\varepsilon$ is lower than that of larger initial $\varepsilon$.}
\label{edg}
\end{figure}

\begin{figure*}
  \begin{center}
    \resizebox{.75\textwidth}{!}{\includegraphics{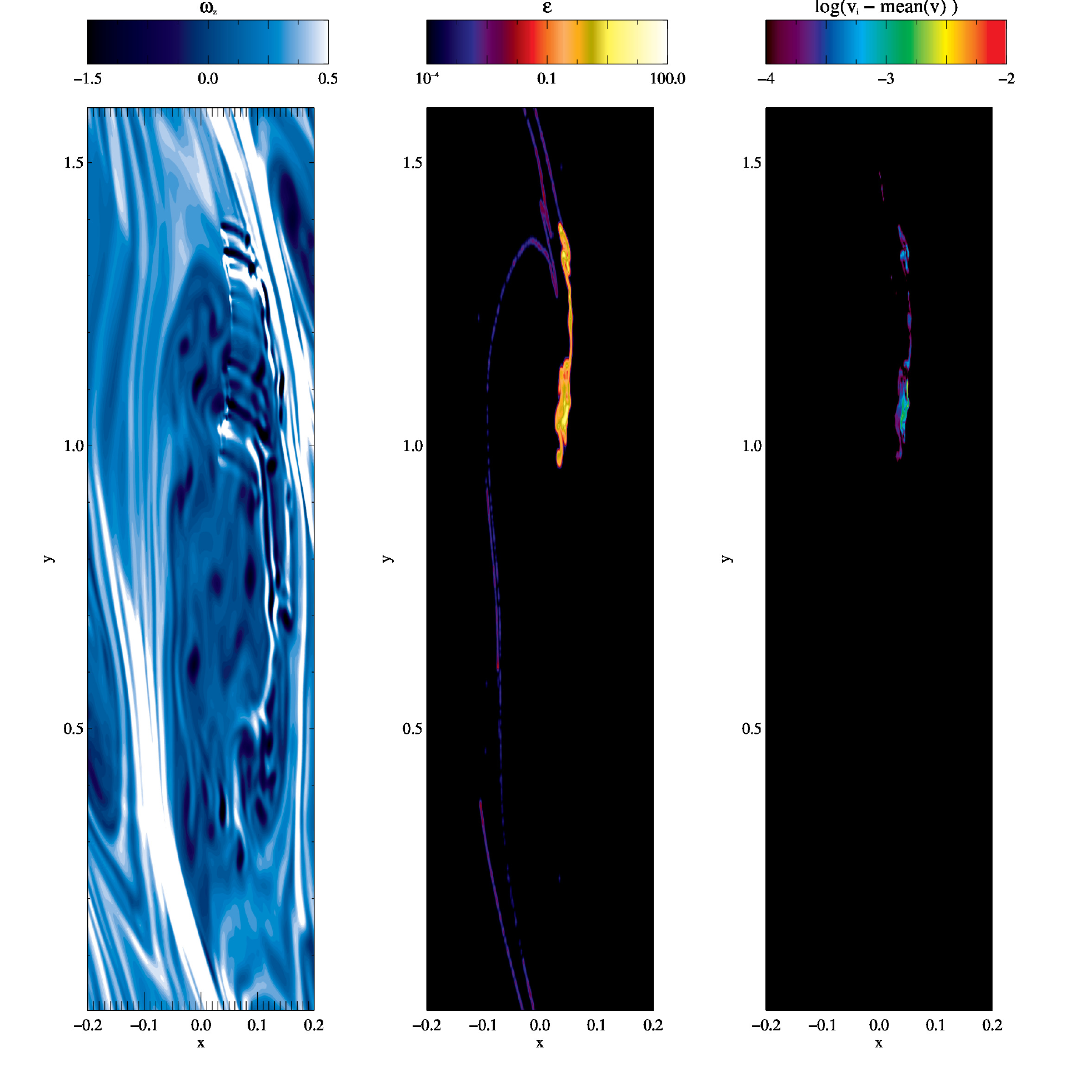}}
    \end{center}
\caption{Vorticity $\omega_z$ in units of the local Keplerian frequency $\varOmega_0$ of the gas flow (left), dust-to-gas-ratio (middle) and particle collisional velocity (right) for St = 1 particles, $\varepsilon_0 = 10^{-2}$ and after 200 local orbits. Collisional velocities are higher where high particle concentrations are located, but hardly exceed $10^{-3}c_s$.}
\label{vcol_1_e-2}
\end{figure*}

\begin{figure*}
  \begin{center}
  \resizebox{.75\textwidth}{!}{\includegraphics{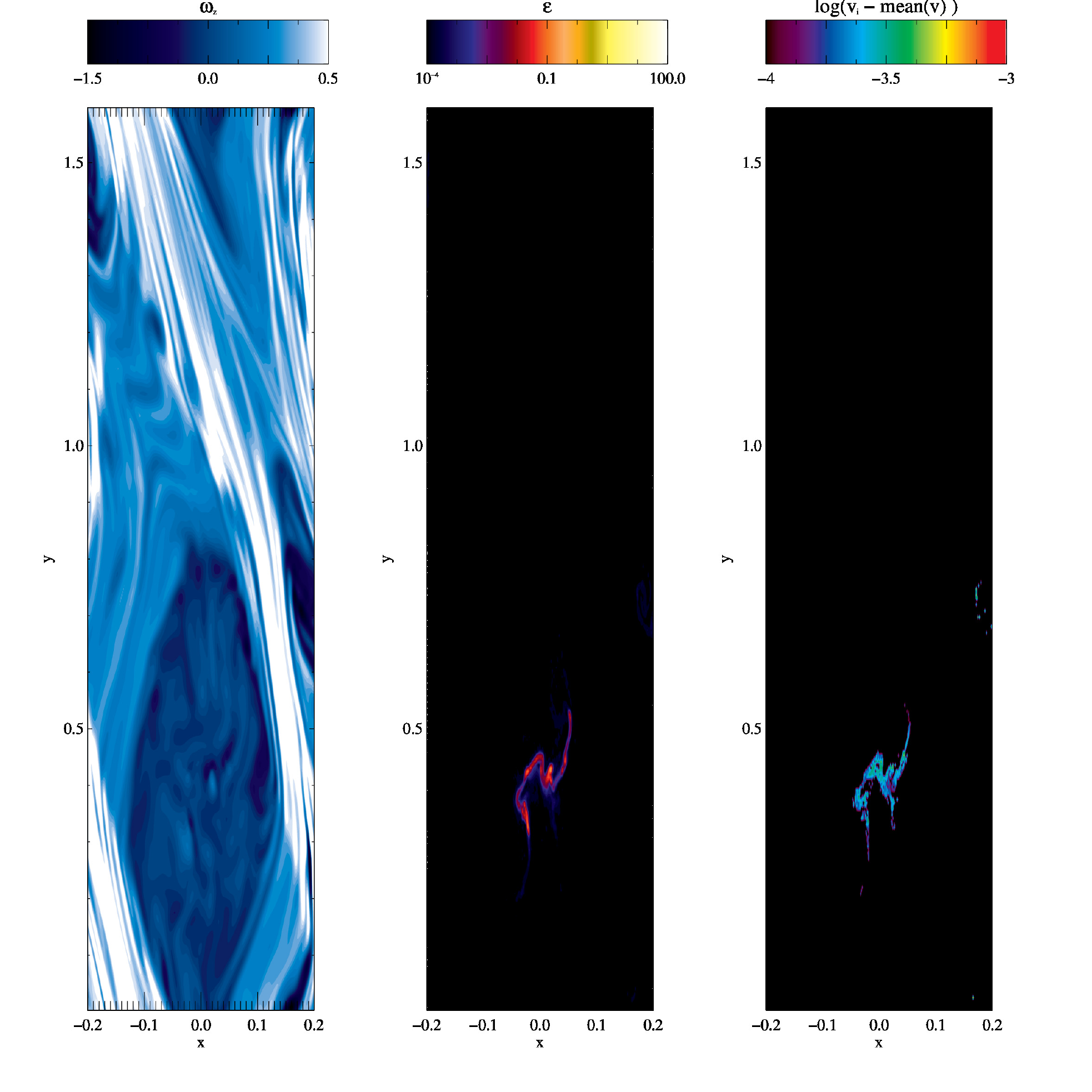}}
  \end{center}
\caption[]{Vorticity $\omega_z$ in units of the local Keplerian frequency $\varOmega_0$ of the gas flow (left), dust-to-gas-ratio (middle) and particle collisional velocity (right) for St = 0.05 particles, $\varepsilon_0 = 10^{-4}$ and after 200 local orbits. Collisional velocities are higher where high particle concentrations are located, but hardly exceed $10^{-4}c_s$.}
\label{vcol_05_e-4}
\end{figure*}

\begin{figure}
\resizebox{\columnwidth}{!}{\includegraphics{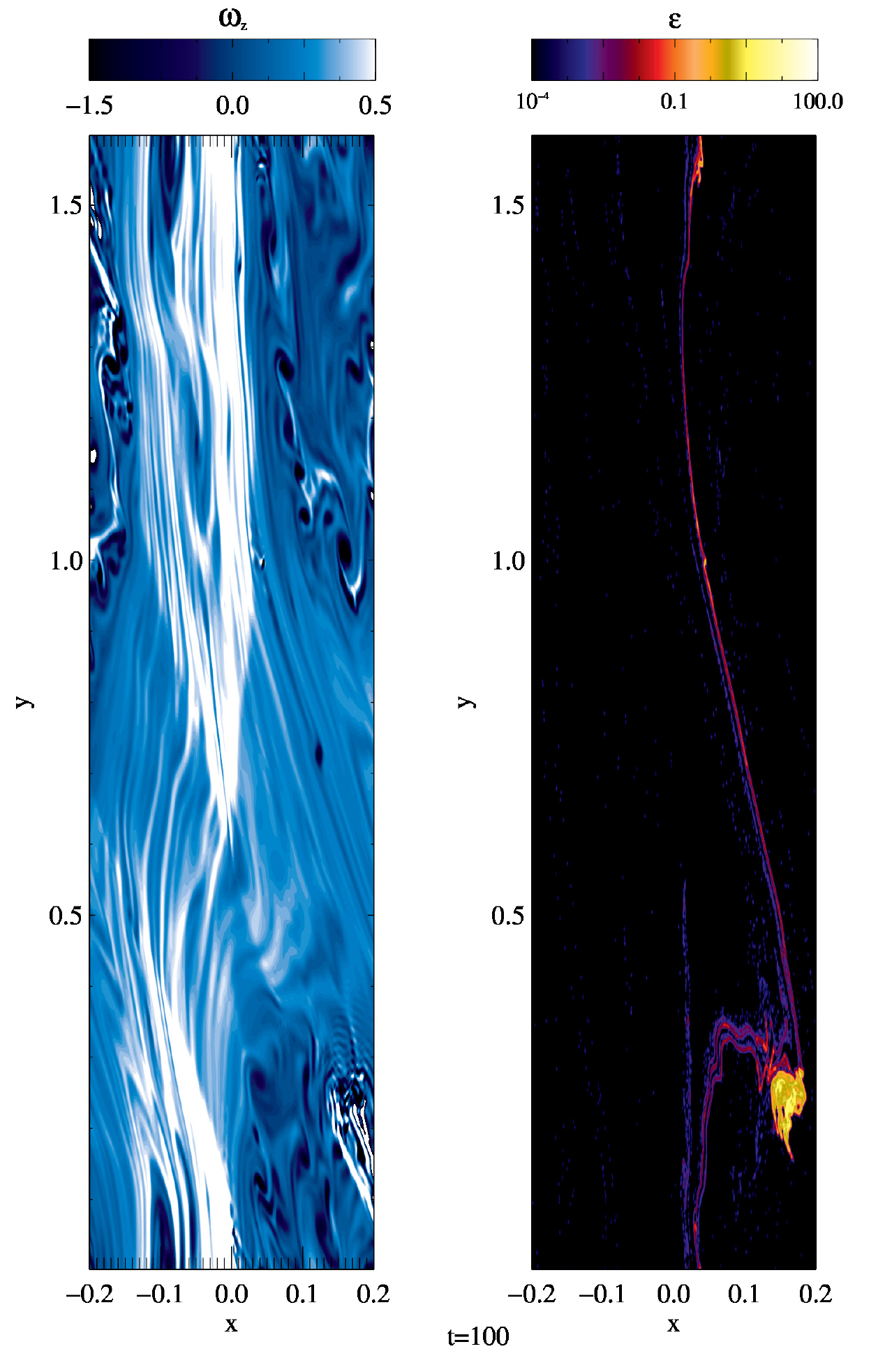}}
\caption{Gas vorticity $\omega_z$ in units of the local Keplerian frequency $\varOmega_0$ and dust-to-gas ratio for St=1 particles with twice the standard resolution. Streaming is still clearly visible and there is no significant difference with respect to the lower resolution simulation (see \Fig{St20u1}).}
\label{high_res}
\end{figure}

\begin{figure}
\begin{center}
\resizebox{.5\columnwidth}{!}{\includegraphics{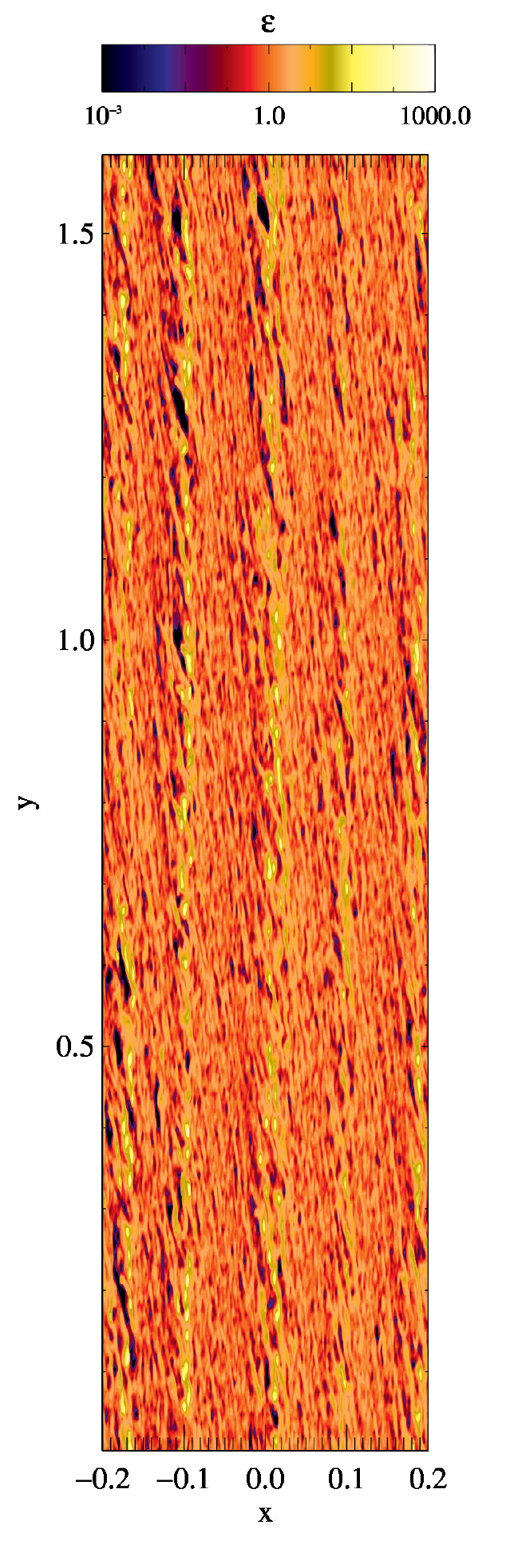}}
\end{center}
\caption{Dust-to-gas ratio for St=1 particles, an initial dust-to-gas ratio of $\varepsilon_0=3$ and no vortex. Clumping of particles still happens, therefore the streaming instability is not an effect of the vortex capturing alone. It can also be resolved with standard resolution and without radial stratification.}
\label{no_vortex}
\end{figure}

\subsection{No Particle Feedback}

We first consider simulations where we do not include back-reaction from the particles, effectively setting $\rho_{\rm d}=0$ in Equation (\ref{NS}). That means that gas drags particles along, but even high dust densities do not affect the gas velocity in any way. 
{The purpose of these simulations is to serve as comparison to the later models in which feedback is included.}

\Figure{edg-nfb} shows the maximum dust-to-gas ratio for these simulations. We see from the almost constant dash-dotted green line that all of the available solid material in the $\rm{St}=1$ particle simulations accumulates in the vortex. This corresponds to a particle concentration for St=1 particles of $c\approx\Sigma_{\rm d}/\Sigma_{\rm d,0}\approx12 \, 000$ and $\varepsilon\approx320$. $\rm{St}=0.05$ eventually also all accumulate within the vortex and then are kept trapped there. The maximum particle concentration is reached sooner for $\rm{St}=1$ particles than for $\rm{St}=0.05$ particles, which can be attributed to their higher radial drift velocity. Even for $\rm{St}=0.01$ particles the dust-to-gas ratio increases to $\varepsilon\approx1$.

We see two different behaviors. For $\rm{St}=0.05$ and $\rm{St}=1$, basically all particles are collected within the vortex, although $\rm{St}=0.05$ take 10 times longer, yet in both cases concentrations of about 4 orders of magnitude are reached. For $\rm{St}=0.01$ concentration inside the vortex is only by 2 orders of magnitude and $\rm{St}=20$ particles concentrate by one order of magnitude, but interestingly outside the vortex in correlation with the zonal flow related to the radial position of the vortex (see \Fig{zonal_flow}). These zonal flows show up in the azimuthally averaged rotation profile, both in magneto hydrodynamical simulations \citep{Lyraetal2008, Johansenetal2009, Dittrichetal2013} and also in our hydrodynamical simulations of vortices. The zonal flows  are deviations from the sub-Keplerian mean rotation profile confined by long-lived radial pressure variations (geostrophic balance).

The different behavior of particle accumulation can be explained through the gas coupling. Particles with $\rm{St}=0.01$ are strongly coupled to the gas. {As an effect it takes much longer to capture the particles into the vortex and as a second effect the deviation between the actual vortex and a analytical solution like for instance a Kida vortex (see discussion in \citealt{LyraLin2013}) prevents the particles from accumulating too densely.} An additional effect to prevent too strong concentrations in 3D vortices might be turbulence, triggered for instance by elliptical instability \citep{LyraLin2013} and internal circulation \citep{Meheutetal2012a}. The $\rm{St}=20$ particles are the other extreme: they are hardly coupled to the gas at all, and therefore are only weakly affected by the vortical motion of the gas. In a stationary analytic vortex these particles would get trapped if one waits long enough. Yet, our numerical vortex is dynamically active in changing its strength and shape and additionally drifting in the radial direction. This vortex dynamics gives the large particles the chance to escape from the vortex again, but still stopped from radial drift in the zonal flow.

\subsection{Including Particle Feedback}\label{PF}

In the last Section we showed that there are significant particle overdensities within the vortex. This means that we cannot neglect the momentum transfer from dust onto gas. Because of this we now include dust back-reaction on the gas, using the full Equation (\ref{NS}).

For the $\rm{St}=20$ particles there is no change with respect to the simulations without particle feedback because overdensities stay below $\varepsilon = 1$ (compare the left and right side of the top panel in \Fig{St20u1}). For these large particles the streaming instability can not be triggered at the present dust-to-gas ratio. All other tested particles still accumulate in the vortex, yet differently:

Particles with $\rm{St}=1$ concentrate more locally (in a smaller area) than smaller particles (bottom panel of \Fig{St20u1} and \Fig{St05u01}). St=1 particles are getting concentrated on time scales which are not longer than the dynamical time scale of the vortex \citep{BargeSommeria1995}, thus they can easily follow the changes of the vortex in shape, strength and location. Smaller particles have much longer time scales to spiral into the center of the vortex - longer than the dynamical time scale of the vortex. As the attractor inside the vortex is changing its location on dynamical time scales (change of amplitude, shape and location of the vortex), the particles have no chance to ever catch up or follow it. 

An effect that occurs when $\varepsilon$ approaches unity is the streaming instability \citep{YoudinGoodman2005, YoudinJohansen2007}. Once particles concentrate, their locally increased dust-to-gas ratio leads to a slower radial drift \citep{Nakagawaetal1986}. This produces a further enhancement of solids since faster particles from slightly larger radii bump into the accumulation, like a traffic jam, which eventually results in streaming dust structures. This is also the case in our simulations, although with our resolution it is possible to resolve only a part of the unstable wavelengths . Therefore we are not able to study the linear and nonlinear behavior of the streaming instability in all detail and model the correct growth-rates of the instability. Yet, for the wavelength we resolve we find the right threshold dust-to-gas ratio to trigger the instability. See also the appendix for additional resolution tests on the streaming instability.

\Figure{St01_sigma} 
shows the vertical gas vorticity $\omega_z$ in units of the local Keplerian frequency $\varOmega_0$ and dust-to-gas ratio $\varepsilon=\Sigma_{\rm g}/\Sigma_{\rm d}$ for a simulation with $\rm{St}=0.01$ particles after 300 local orbits (run F1). It is clear that the particles accumulate inside of the vortex and follow the vortical motion. Where the concentration is highest they create strong maxima in the gas vorticity. This is an effect of the cyclonic rotation of the particle clumps which was already reported in earlier work \citep{LambrechtsJohansen2012} and is a consequence of the conservation of angular momentum under contraction which can turn anticyclonic motion in a rotating system to cyclonic motion. 

In general due to the back-reaction of the dust on the gas and the resulting streaming instability the initially elliptical gas streamlines are bent into more complex motions than in cases without back-reaction.

In case of the $\rm{St}=1$ particles where the local particle concentration, and therefore the back-reaction, are strongest, the vortex structure is disrupted. Because of this, the particle trapping mechanisms lose strength: pressure gradients across the vortex become shallower, Coriolis forces in the vortex grow weaker, and the particles begin to escape the vortex. Because the local particle concentration decreases, the large vorticity gradients flatten out again. This eventually leads to a new amplification of the vortex due to the background stratification and the process repeats itself (see series of snapshots in \Fig{snapshots} of the electronic version of this paper).

{The destruction of the vortex occurs on very short time scales set by the streaming instability, whose growth time is a function of the dust to gas ratio and Stokes number of the dust that can be as fast as one orbital period \citep{JohansenYoudin2007}. The new vortex amplification on the other hand takes hundreds of orbits like in the initial growth as it depends on the radial stratification of the disk and the thermal relaxation time of gas.}

This vortex disruption and reforming process discussed here is not to be confused with the vortex instability discussed in \citet{ChangOishi2010}. They analyze the stability of a three dimensional vortex based on the density contrast between the interior of a vortex and the ambient medium and determine that if this contrast is higher than a few 10\% then the vortex will become unstable. Although we see such high dust density concentrations inside the vortex they are very localized and therefore our vortices remain stable. For them to become entirely unstable according to Chang and Oishi's analysis the density contrast needs to be uniformly high, spread out over the entire vortex, which is never the case in our simulations. 

\Figure{edg_fb} shows the maximum dust-to-gas ratio of our different simulations with different ${\rm St}$. We clearly see that $\rm{St}=1$ particles (dash-dotted green line) have the highest concentration. As the particle size decreases, the concentration also decreases. It is important to note that $\rm{St}=20$ particles do not accumulate inside the vortex. The highest dust-to gas ratio is reached for $\rm{St}=1$ particles. Larger and lower ${\rm St}$ reach lower dust-to-gas ratios.

We now turn our attention to the spatial distribution of the dust concentration. By clustering we understand what fraction of the dust takes part in the high overdensities which are the particles triggering the streaming instability. In \Fig{clustering} we show what fraction of the entire dust content $M_{\rm d}$ has a specific dust-to-gas ratio. The dashed vertical lines indicate the initial dust-to-gas ratio $\varepsilon_0=0.01$ and $\varepsilon=1$. For e.g. $\rm{St}=1$ particles 83.74\% have accumulated in regions with $\varepsilon\ge 1$ whereas for $\rm{St}=0.05$ particles only 2.13\% of the entire dust mass is concentrated in areas with $\varepsilon\ge 1$. The remaining dust is spread out thinner. All results can be see in the 7th and 8th column of Table \ref{setup}. Note that for the clustering we considered a temporal average, but for the maximum dust-to-gas ratio $\varepsilon_{\rm max}$ (5th column in Table \ref{setup}) we took the absolute maximum form the entire run. Therefore for runs F1 and DG1 $\varepsilon_{\rm max}> 1$ at a given moment while on a temporal average this value is not reached. A dust-to-gas ratio of $\varepsilon=1$ is significant, because that is when dust back-reaction on the gas becomes important. This means that for $\rm{St}=1$ back-reaction is a requirement if we want to model particle behavior realistically. However, for larger ${\rm St}$ it seems that back-reaction, and with that the streaming instability, contributes little to the overall dynamical behavior of particles. Yet, we already saw that there is a significant difference between the simulations with and without particle feedback for $\rm{St}=0.05$ particles (see \Figs{edg-nfb} {edg_fb}). Without feedback all particles were accumulated in the vortex whereas the maximum dust-to-gas ratio with feedback was around $2 - 8$, two orders of magnitude lower.This is confirmed by the \Fig{compare}, where we compare the clustering of St=0.05 particles with (dashed line) and without feedback (solid line).  The particle accumulation is similar for low dust-to-gas ratios. However, as $\varepsilon$ increases, streaming sets in for the simulation with feedback and thus regulates the overdensities. Without feedback the overdensities can grow unhinderedly which results in about 80\% of all particles accumulating in $\varepsilon> 1$. We want to stress that although the dust-to-gas ratios are different depending on whether back-reaction is included or not, it might be difficult to distinguish such extremely large local confined concentrations observationally with e.g. ALMA \citep{LyraLin2013}. Also for $\rm{St}=0.01$ particles that only reach $\varepsilon_{\rm max}\approx 1$ we already saw that there is an effect on the gas (see \Fig{St01_sigma}). In this context one has to stress that growth rates and unstable wavelengths towards the streaming instability depend on the St as well as on the actual dust-to-gas ratio. A detailed study of the non linear evolution of the streaming instability and the clumping caused by the streaming instability can be found in \citet{Johansenetal2009} who can afford a resolution 10 times higher than we do. One can interpret the simulations of  \citet{Johansenetal2009} as a detailed study on the dust behavior inside a vortex concentration and therefore our vortices generate the dust-to-gas overdensities needed for the  \citet{Johansenetal2009} scenario.

 We conclude that as soon as even a fraction of particles approaches $\varepsilon \approx 1$ back-reaction from the particles onto the gas needs to be included to accurately model their behavior.

\subsection{Lower Dust-to-Gas ratios}

So far the considered initial dust-to-gas ratio always was $\varepsilon_0=1:100$, but it has been shown that planets can also form in low metallicity disks \citep{Mordasinietal2012, Niedzielskietal2009}. It is also possible that planetesimal formation is still ongoing in a disk where some planets have already formed like in the 51 Peg system \citep{Dumusqueetal2012} and therefore the disk can be depleted in solids. A third reason to consider lower initial dust-to-gas ratios is the possibility that only a small fraction of the entire dust content is in the particle size regime that we study as a result of the coagulation-fragmentation balance as studied by \citet{Birnstieletal2012}. To account for this we perform simulations with lower initial dust-to-gas ratios like $\varepsilon_0=1:1\, 000$ and $\varepsilon_0=1:10\, 000$. The number of super-particles we put into the domain stays the same, while each super-particle represents less mass than in cases with higher initial dust-to-gas ratio.

\Figure{edg} shows the particle accumulation and resulting dust-to-gas ratios for $\rm{St}=0.05$ (simulations F2, DG1, DG2) and $\rm{St}=1$ (simulations F3, DG3, DG4). In cases with low $\varepsilon_0$ more super-particles concentrate in one location. Since each of these super-particles has less mass compared to in simulations with higher dust-to-gas ratio, the back-reaction is less effective. Thus the absolute dust-to-gas ratio values to be reached in low metallicity systems are almost as high as in high metallicity systems. The gas is not affected as much by the dust particles as in previous simulations. Thus the vortices, although still disrupted slightly by the back-reaction, are no longer torn apart. The particles are trapped more tightly and cannot leave the vortex.

We conclude that the relative concentration in low metallicity disks is much stronger than in high metallicity systems, thus the absolute dust-to-gas ratio values to be reached in low metallicity systems is almost as high as in high metallicity systems. In the end always a roughly identical dust-to-gas ratio is reached (within one order of magnitude \r{edg}), e.g. at the physical condition for triggering the streaming instability.

\section{Collisional Velocities}

In our simulations collisions between the particles are neglected. Already for the physical conditions of a protoplanetary disk with real particle numbers a typical collision time, i.e. the time after all particles have collided once is typically a few thousand orbits. With our super-particle approach we have no chance to ever observe a real collision in the course of our simulation. Still collision speeds are important to know to study possible growth or disruption in statistical codes \citep[e.g.][]{Birnstieletal2012}.

So when we talk about collisional velocities we rather mean relative velocities of neighboring particles from which we extrapolate likely collision speeds.
In the following we will discuss how we calculate their collisional velocities.

Since we use a two-dimensional approach, our model can only provide first estimates for collisional velocities. For instance we neglect small scale turbulence that will arise from the vertical 
structure of the disk and particle layer \citep[e.g. Kelvin-Helmholz: see][]{Johansenetal2006b} and elliptic instability due to 3D waves propagating inside the vortex,  but also from the tail of the Kolmogorrov cascade that is not present in 2D simulations. In general 2D flows tend to create large vortical structures rather than allowing energy to cascade down to the dissipation scale.

Therefore we take a rather simple approach to estimate collisional velocities, which are basically the local dispersion in the particle velocities. More exactly: We consider each grid cell separately, since only particles which are close together have a chance to collide. From each individual particle's velocity $\v{v_i}$ in a given grid cell the mean particle velocity of all particles within this  grid cell $\v{\bar v}$ is subtracted. Afterwards the average of these residual velocities within one grid cell is taken. This results in the mean collisional velocity within this grid cell. To get the average collisional velocity of the entire system $v_{\rm coll}$ we average over all grid cells that include at least 10 particles. The results can be seen in Table \ref{v_coll} for St=1 and St=0.05 particles. \Figs{vcol_1_e-2}{vcol_05_e-4} show the spacial distribution of the collisional velocities. Higher collisional velocities coincide with regions of high particle density.

For tightly coupled particles (low Stokes numbers) it is neccessary to first subtract the local gas velocity at the position of a particle from $\v{v_i}$. Otherwise we would measure the velocity dispersion of the gas rather than of the particles. This results in low collisional velocities $\Delta v_{\rm coll}$, for the smaller St=0.05 particles. This is not so much the case for St=1 particles as they only marginally couple to the gas.

If we compare these collisional velocities to the theoretically predicted collisional velocities by \citet{OrmelCuzzi2007} for a given turbulent $\alpha$ value we see grave deviations. Already our gas velocity $u_{\rm rms}$, which is 3 to 4 times as large as the turbulent gas velocity $u_{t} = \sqrt{\alpha} c_s$, where $\alpha$ is a dimensionless viscosity parameter \citep{ShakuraSunyaev1973}, suggest a deviation from their model. For St = 1 particles they predict a ratio of particle collisional velocities to turbulent velocities of 1, yet our value is of the order of $10^{-3}$. This shows that the theory of Kolmogorov turbulence is not applicable to the vortex environment, at least as long as we neglect the third dimension and thus cannot predict particle collisions inside vortices accurately.

From our model we see that the $\alpha$ values measured outside the vortex are not a good proxy to estimate the collisional velocities inside the vortex. In our 2D simulations the vortex may appear calmer than it would actually be in 3D where elliptical instability can produce turbulence inside the vortex core \citep{LesurPapaloizou2010, LyraKlahr2011}.

Yet, as we stated before, this is a rather simple approach and these collisional velocities can only be seen as a first estimate. There are several factors that can increase the velocities. Higher resolution and a treatment of three dimensions will lead to properly resolving the streaming instability and with this particle velocities will increase \citep{Johansenetal2009}. Including the vertical stratification will increase particle velocities, too, as also the vortex dynamics become more complex.

{It is not possible to quantify the error introduced by our two-dimensional approximation. For that purpose we have to wait until we have evaluated full three-dimensional models, which are numerical wise significantly more expensive.}

\section{Particles Collected inside the Vortex and Possible Gravitational Collapse}

Our 2D simulations tell us that the dust-to-gas ratio is increased significantly inside a vortex.
A detailed study on gravitational fragmentation of the particle layer would require 3D modeling as the criterion for collapse is to reach a critical 3D density of $\rho_{\rm R}$, which is the Roche density. The Roche density is defined as the density of a clump that cannot be sheared apart by tidal forces from the central object. 

It can be derived by equating the the gravitational acceleration on the surface of a clump
\begin{equation}
a_{g,\rm c}=\frac{GM_{\rm c}}{R_{\rm c}^2},
\end{equation}
where $M_{\rm c}$ and $R_{\rm c}$ are the clumps mass and radius respectively, with the tidal acceleration $a_t$ on the clumps surface. The tidal acceleration is the difference between the gravitational pull of the central star at the clumps surface and its center of mass. To first order we get
\begin{equation}
a_{t}\approx\mp2R_{\rm c}\frac{GM_\star}{r^3},
\end{equation}
where $M_\star$ is the mass of the central object and $r$ is the distance between star and particle clump. If we now equate these two accelerations, exchange the masses by their respective densities and solve for the clump density we get
\begin{equation}
\rho_{\rm crit}> \rho_{\rm R}=\frac{3}{2}\frac{M_\star}{\pi r^3}.
\end{equation}
If the density of the particle clump is higher than this critical density it will be held together by its own gravity. If this is not the case then the particle clump will be torn apart by tidal forces.

There are additional forces that act to prevent collapse besides gravitational tides, e.g. erosion and tides exerted by the vortex \citep{Lyraetal2009b}, but we only aim for a first estimate here and therefore neglect these forces.

If we take the mass of the Sun for the mass of the central object then we get $\rho_{\rm R}(1\,{\rm AU})=2.83\xtimes{}{-7}\,{\rm g \, cm ^{-3}}$ and $\rho_{\rm R}(5.2\, {\rm AU})=\xtimes{2.01}{-9}{\,\rm g\, cm ^{-3}}$. The highest overdensities we observed reached for $\rm{St}=1$ particles is $\Sigma_{\rm d}\approx80\Sigma_{\rm g}$. This means that the trapping alone will in most cases not lead to Roche densities, yet we have neglected sedimentation. As was shown in \citet{Johansenetal2009} an increase in the metallicity of the disk by a factor of a few is sufficient to trigger both streaming instability and in succession gravitational collapse into planetesimals.

For St=0.01 particles for example it is apparent from \Fig{St01_sigma} that the material inside the vortex structure reaches dust-to-gas ratios of $\varepsilon\approx0.1$. This is a concentration in local available dust by a factor of 10. As shown by  \citet{Johansenetal2009} a concentration by a factor of 3 would be sufficient to trigger planetesimal formation. 

Nowhere outside the vortex these $\varepsilon$ values are reached. Comparing this to the clustering in \Fig{clustering} we see that about 40\% of the entire particle fraction are in concentrations of $\varepsilon\ge 0.1$ and thus about 40\% of the entire particle fraction is captured inside the vortex. Doing the same analysis for the other particle sizes even leads to values up to 98\% which means that the capturing mechanism is very efficient. Note that these are best case values, since in our simulation no material is lost. If a particle is not captured by the vortex the first time it passes it, it can be captured the second time due to the periodic boundaries, whereas in a real disk this particle would be lost to the vortex. On the other hand we also neglect the influx of particles from larger radii \citep[see][]{Birnstieletal2012}, thus one can argue that our fixed global metallicity is a conservative under estimation, as even more material might end up in the vortical trapping region.

\section{Summary and Conclusion}\label{2Dconclusion}

In this paper we analyze how particles and vortices sustained by the radial stratification of a disk affect each other. In particular, we investigated whether particles of various sizes can concentrate in vortices in the presence of particle feedback, and whether vortices remain a long-lived phenomenon in this case. 

We have conducted two sets of simulations for a series of Stokes numbers ($\rm{St}=20, 1, 0.05$ and $0.01$): one where there is only gas drag on the particles and one where particles also exert drag on gas. This becomes important if the initial dust-to-gas ratio is locally enhanced from 0.01 to about 1. 

We see that, without back-reaction, particles of $\rm{St}=0.05$, and $\rm{St}=1$ are swept up very efficiently by the vortices. For St=1 the concentration is so efficient that all particles end up in a single grid cell. Smaller and larger particles can escape from the vortex again. For smaller particles this is because they are too slow to follow the dynamical evolution of the vortices. Larger particles leave the vortex in the azimuthal direction, yet get trapped in the zonal flow correlated with the radial position of the vortex.

The dynamics changes considerably when particle feedback is included. $\rm{St}=20$ particles are hardly affected by the vortex structure and no critical particle concentration to trigger the streaming instability is reached. Therefore the gas (and thus the vortex) is not affected by these large particles. In conclusion, if smaller particles coagulate to sizes corresponding to St=20, these bodies would neither be captured by vortices nor would they be retained inside them. Yet we highlight that our simulations show these bodies of large Stokes number getting trapped in the zonal flow related to the vortex. This effect has already been observed in global simulations \citep[Figure 11 in][]{Lyraetal2009b}.

For smaller sizes we see high concentrations inside the vortex. Particles of $\rm{St}=1$ concentrate very efficiently in the center of the vortex, leading to streaming instability. The instability leads to strong vorticity perturbations, that eventually disrupt the vortex. With decreasing particle density the vortex can be re-established and the cycle repeats. The dust-to-gas ratio can be locally increased up to $\varepsilon\approx 80$. We measure that over 80\% of the dust is concentrated in grid cells with $\varepsilon >1$.

Smaller particles are not that strongly concentrated. Instead they are spread out over the entire vortex and take part in the streaming instability \citep{YoudinGoodman2005, YoudinJohansen2007}. This instability typically shows up when  $\varepsilon > 1$ even when only a small fraction of the available dust takes part in the instability. For $\rm{St}=0.05$ particles  only 2.13\% of the entire mass reached $\varepsilon > 1$. The maximum dust-to-gas ratio reached are $\varepsilon\approx 3$ for $\rm{St}=0.05$ and $\varepsilon\approx 1$ for $\rm{St}=0.01$.

Although the particle concentrations achieved are quite different for the different particle sizes, the overall mass of particles accumulated in a vortex is roughly the same which matches quite well the analytical prediction of \citet{LyraLin2013}. Around 90\% of the entire dust content is swept up by the vortex. This corresponds to a dust density about four times higher inside the vortex than in the background, and about 30 times higher than in the region outside the vortex. This highlights the prospects of detecting vortices by their enhanced dust to gas ratio \citep{WolfKlahr2002,vanderMareletal2013}.

We also conducted simulations with lower initial dust-to-gas ratios, for three reasons. These are: 1) to see whether planetesimal formation can be triggered around low metallicity stars; 2) to see if evolved dust populations in disks (with significant dust mass in at least planetesimal-mass objects or already accreted due to radial drift) can still form planetesimals, and; 3) to mimic a situation in which only a small fraction of the total dust content grows to a size of ${\rm St=0.01} - {\rm St=1}$ as a result of the coagulation-bouncing-fragmentation-drift balance \citep{Birnstieletal2012}. We see that although the initial dust-to-gas ratio $\varepsilon_0$ is a factor 10 or 100 lower, the locally reached maximal dust-to-gas ratio is still of the same order of magnitude. The relative particle concentration increases until the back-reaction becomes significant, which is independent of the initial dust-to-gas ratio.

The first estimate of the collisional velocities of the dust particles are much lower than what is to be expected based on Kolmogorov turbulence of equivalent $\alpha$. 
{These low velocities are possibly a result of our two-dimensional approximation and will have to be tested in three-dimensional simulations to accurately calculate collision speeds and its effect of limiting particle accumulation.}

We confirm that baroclinic vortices are an important mechanism for accumulating particles even when feedback via drag is considered. The concentrations achieved are, depending on particle size, a factor $100$ to $10\,000$ higher than the average value. Even if there is only a low amount of dust present, these high overdensities can be reached. Streaming instability additionally enhances the dust concentration. The present 2D study neglects the vertical structure of the gas disk and the dust layer. The expected difference in 3D vs. 2D simulations lies in the additional concentration effect by vertical sedimentation that is obviously not possible in 2D vertically integrated models. Thus, 2D studies can be said to conservatively underestimate the maximum particle concentration that can be reached in 3D. On the other hand, in three dimensions, the vortex itself is weaker due to elliptic instability \citep{LesurPapaloizou2009, LyraKlahr2011}. Also, due to this instability, enhanced particle diffusion is acting inside the vortex. These issues will be addressed in a future 3D study. 


\appendix

\section{Test of Streaming Instability}

As we stated in Section \ref{PF} we expect the streaming instability to occur due to the achieved dust densities of $\varepsilon \geq 1$. However, with the used resolution we do not resolve all possible modes. We performed two tests in order to ascertain that the streaming instability is indeed present. First we simply doubled the numerical resolution, so that smaller streaming modes are resolved. For the second test, we modeled a box without a vortex or radial stratification ($\beta = 0$), but increased the initial dust-to-gas ratio to a value that would trigger the streaming instability ($\varepsilon=3$). 

There was no significant difference between our standard resolution and the first test case (\Figs{high_res}{St20u1}, lower right). We see the same streaming structures in the particle density, and the achieved overdensities in both cases are comparable. Therefore we can deduce that the standard resolution we used for all simulations is high enough to capture the essence of the streaming instability.

The second case also showed streaming structures (\Fig{no_vortex}). This shows that our resolution is high enough to enable the streaming instability, and that the streaming structures are not simply a result of particle accumulation in the vortex. The vortex accumulation first produces a high concentration of particles, that is 
in turn amplified by the streaming instability, further increasing the local dust-to-gas ratio.


\end{document}